\documentclass[useAMS,usenatbib]{mn2e}
\usepackage[percent]{overpic}
\usepackage{times}
\usepackage{graphicx}
\usepackage{amsmath}
\usepackage{mathabx}
\usepackage{subfigure}
\usepackage{natbib}
\usepackage{txfonts}
\usepackage{journals}
\usepackage{xcolor}

\newcommand{\ks}{$K_{\rm S}$}
\newcommand{\gb}{$G_{\rm BP}$}
\newcommand{\gr}{$G_{\rm RP}$}
\newcommand{\egk}{$E(G-K_{\rm S})$}
\newcommand{\ebr}{$E(G_{\rm BP}-G_{\rm RP})$}
\newcommand{\ehk}{$E(H-K_{\rm S})$}

\title[3D Reddening maps in disk]
{Three-dimensional interstellar dust reddening maps 
of the Galactic plane} 

\author[B.Q. Chen et al.]
{B.-Q. Chen,$^{1}$\thanks{E-mail:
bchen@ynu.edu.cn (BQC); x.liu@ynu.edu.cn (XWL).}
 Y. Huang,$^{1}$
 H.-B. Yuan,$^2$
 C. Wang,$^3$
  D.-W. Fan,$^4$
 M.-S. Xiang,$^4$
   \newauthor
 H.-W. Zhang,$^3$
 Z.-J. Tian,$^{3,5}$
and  X.-W. Liu$^{1}$\footnotemark[1]
\\
$^{1}$South-Western Institute for Astronomy Research, Yunnan University, Kunming, Yunnan 650091, P.\,R.\,China\\
$^{2}$Department of Astronomy, Beijing Normal University, Beijing 100875, P.\,R.\,China\\
$^{3}$Department of Astronomy, Peking University, Beijing 100871, P.\,R.\,China\\
$^{4}$National Astronomy Observatories, Chinese Academy of Sciences, Beijing 100101, P.\,R.\,China \\
$^{5}$Department of Astronomy, Yunnan University, Kunming, Yunnan 650091, P.\,R.\,China\\
 }

\begin{document}

\date{Accepted ???. Received ???; in original form ???}

\pagerange{\pageref{firstpage}--\pageref{lastpage}} \pubyear{2016}
\maketitle
\label{firstpage}

\begin{abstract}
We present new three-dimensional (3D) interstellar dust reddening maps
of the Galactic plane in three colours, \egk, \ebr\ and \ehk.
The maps have a spatial angular resolution of 6\,arcmin and covers 
over 7000\,deg$^2$ of the Galactic plane for Galactic longitude 
0\degr\ $<$ $l$ $<$ 360\degr\ and latitude $|b|$ $<$ $10$\degr.
The maps are constructed from robust parallax estimates from the 
Gaia Data Release 2 (Gaia DR2) combined with the 
high-quality optical photometry from the Gaia DR2
 and the infrared photometry from the 2MASS and WISE surveys.
We estimate the colour excesses,  \egk, \ebr\ and \ehk, of over 56 million stars with the
machine learning algorithm Random Forest regression, using a training data set
constructed from the large-scale spectroscopic surveys LAMOST, SEGUE and APOGEE.
The results reveal the large-scale dust distribution in the Galactic disk, showing a number of features 
consistent with the earlier studies. The Galactic dust disk is clearly warped and show complex structures 
possibly spatially associated with the Sagittarius, Local and Perseus arms.
 We also provide 
the empirical extinction coefficients for the Gaia photometry that can be used to convert the 
colour excesses presented here to the line-of-sight extinction values in the Gaia photometric bands.
\end{abstract}


\section{Introduction}

The interstellar dust extinction poses a serious challenge for the study of the structure, 
formation and evolution of our Galaxy,
especially for the low Galactic latitude disk regions \citep{Chen2017a}. 
For any stellar study near the Galactic plane, one needs to correct
for the effects of dust extinction and reddening to interpret the observations properly.

Traditional two-dimensional (2D) extinction maps give the total amount of extinction
in a given direction integrated along the 
line-of-sight. Consequently, the value represents an upper
limit of the real one for a local disc star in that direction. 
In the past decades, thanks to a number of large-scale
photometric and spectroscopic surveys, such as  
the Sloan Digital Sky Survey (SDSS; \citealt{York2000}), the Two Micron All Sky Survey
(2MASS; \citealt{Skrutskie2006}), the LAMOST Experiment for Galactic Understanding and Exploration
(LEGUE; Deng et al. 2012; Zhao et al. 2012) and the Pan-STARRS\,1 Survey (PS1; \citealt{Chambers2016}), 
three-dimensional (3D) extinction maps constructed based on estimates of the distances and  
extinction values to millions of individual stars have become available.
 
Based on the 2MASS data, \citet{Marshall2006} present a three-dimensional (3D) extinction
model of the inner Galaxy ($|l|$ $<$ 100\degr\ and $|b|$ $<$ 10\degr) by
comparing the observed colours of giant stars for each line of sight with the synthetic values
from the Besan\c{c}on Galactic model \citep{Robin2003}. Using a similar method, 
\citet{Chen2013} and \citet{Schultheis2014} present the colour excess maps for 
several colours toward the Galactic Bulge ($|l|$ $<$ 10\degr\ and 
$-$10\degr\ $<$ $b$ $<$ 5\degr) based on data from 
the ESO Public Survey, VISTA Variables in the Via Lactea (VVV; 
\citealt{Minniti2010}) and the 
Galactic Legacy Infrared Mid-Plane Survey Experiment
(GLIMPSE; \citealt{Churchwell2009}).
By combining the SDSS optical and the 2MASS near-IR photometry, \citet{Berry2012} simultaneously
estimate distances and values of extinction of stars by
fitting the observed optical (and IR) spectral energy distributions
(SEDs).  Analogously, \citet{Chen2014} present a 3D extinction map that covers 
the entire Xuyi Schmidt Telescope Photometric Survey of the Galactic Anticentre (XSTPS-GAC) survey area
of over 6000\,deg$^2$ (140\degr\ $<$ $l$ $<$ 220\degr and $b$ $<$ 40\degr),
based on the XSTPS-GAC optical and the 2MASS and Wide-Field Infrared Survey 
Explorer (WISE; \citealt{Wright2010}) IR photometry. 
Using a hierarchical Bayesian model, \citet{Sale2014} derive a 3D extinction map in the Northern Galactic Plane 
(30\degr\ $<$ $l$ $<$ 215\degr and $|b|$ $<$ 5\degr) from the optical 
photometry of the INT/WFC Photometric H$\alpha$ Survey (IPHAS; \citealt{Drew2005}). 
\citet{Green2015} and \citet{Green2018}
apply the Bayesian approach to the PS1 optical and the 2MASS IR photometry to 
produce 3D maps of dust reddening over three quarters of the sky ($\delta$ $>$ $-$30\degr).
\cite{Hanson2016} present 3D extinction maps within a few degrees of the Galactic plane
(0\degr\ $<$ $l$ $<$ 250\degr and from $|b|$ $<$ 4.5\degr) based on
photometry of the PS1 and GLIMPSE surveys. \citet{Gontcharov2017} present a 3D dust reddening map
with a radius of 1200\,pc around the Sun and within 600\,pc of the Galactic midplane based on the colour-magnitude
diagram of the 2MASS photometry. \citet{Lallement2014} and \citet{Lallement2018} apply an inversion method to
data collected from a variety of catalogues and construct 3D extinction
maps of the local interstellar dust.
\citet{Kh2017} and \citet{Kh2018} 
 also developed a 3D inversion method, treating the dust density field as a
Gaussian process.

Distance is a key parameter to build a 3D extinction map. 
Most of the works mentioned above derive
distances to individual stars based on the photometric parallax relations, 
with typical uncertainties of about 20\,per\,cent for dwarfs and more for giants
\citep{Chen2014, Green2014}. 
The second Gaia data release (Gaia DR2, \citealt{Gaia2016, Gaia2018}) contains parallax estimates for 
over one billion stars.  For stars of magnitudes ranging between 8 and 18\,mag, the 
parallax uncertainties vary between $\sim$ 0.04 and 0.1\,mas, corresponding to
distance uncertainties better than  
20\,per\,cent out to $\sim$ 5\,kpc \citep{Lindegren2018}.
Based on data published in the Gaia DR2,
\citet{Bailer2018} provide distance estimates of individual stars 
inferred from procedure that accounts for the non-linearity of the transformation 
and the asymmetry of the resulting probability distribution.
The Gaia DR2 also includes high quality photometry in 
three bands, $G$, \gb\ and \gr\ \citep{Evans2018}. The robust distance
estimates and the precise photometry of Gaia DR2 provide us a great 
opportunity to construct high-quality 3D extinction map of the Galactic disk.

 \citet{Andrae2018} have recently estimated values of extinction $A_G$ and colour excess \ebr\
for 88 million stars based on the photometry and parallax of Gaia DR2. 
However, as they point out, 
their results, using only three optical bands and parallax, suffer from large uncertainties.
There is a significant degeneracy between the intrinsic colours
(or effective temperature) and extinction using just optical colours \citep{Bailer2011, Andrae2018}.
\citet{Berry2012} show that the
degeneracy can be broken (or reduced) by introducing the IR colours.
In the current work, we combine the Gaia optical 
with the 2MASS and WISE IR photometry, and obtain 
robust estimates of colour excess in several colours for tens of millions of stars in the Galactic disk ($|b|$ $<$ 
10\degr). We adopt the Random Forest regression, a machine learning algorithm, to derive the values 
of colour excess. Compared to the traditional SED fitting or 
Bayesian approaches, this machine learning algorithm works much faster to 
provide results of similar accuracies (see Sect.~5.2).
The results are then used to construct 3D 
colour excess maps of the Galactic disk.

The paper is structured as following. In Section 2, we present the
relevant Gaia DR2, 2MASS and WISE data. Section 3 describes
the methods used to derive values of colour excess and to construct 
the 3D colour excess maps.
In Section 4, we present our main results which are
discussed in Section 5. We summarize in Section 6.

\section{Data}

 \begin{figure}
    \centering
  \includegraphics[width=0.48\textwidth]{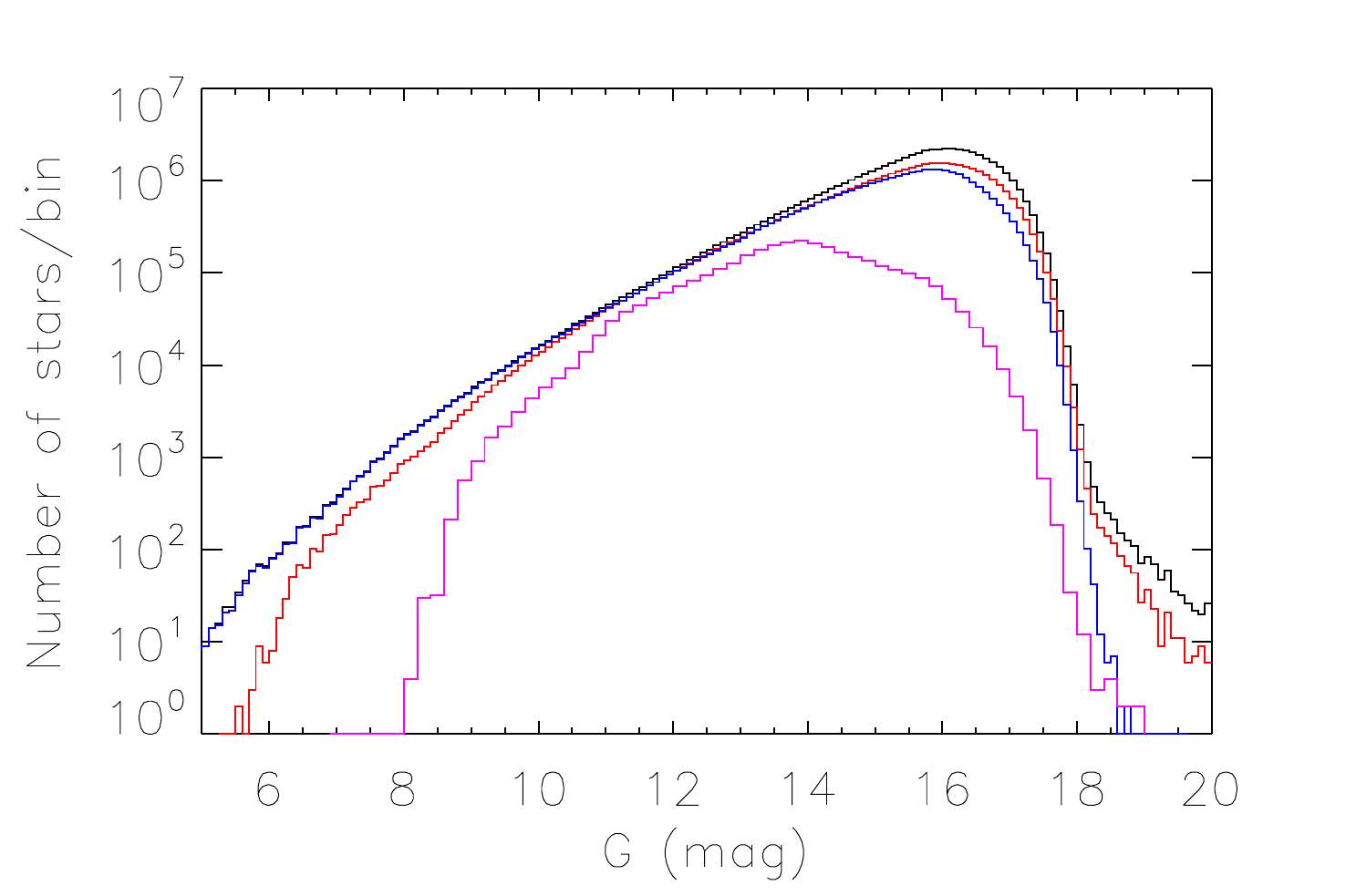}
  \caption{Distributions of  star count as a function of
Gaia $G$-band magnitude for four samples.   The black line represents all over 56 million sources
  in the combined photometric catalogue, whereas those represented by red and blue lines 
  are respectively sources that have WISE $W1$ photometry ($\sim$ 42 million stars ) and that have Gaia parallax 
  uncertainties smaller than 20\,per\,cent ($\sim$ 35 million stars). Finally, the pink line represents sources
  in the training sample ($\sim$ 3 million stars).}
  \label{datamd}
\end{figure}

Our work is based on broadband photometry from three all-sky surveys, the Gaia, 2MASS and WISE.

The Gaia DR2 \citep{Gaia2018}, released in April 2018, provides high precision positions and 
$G$ broad band photometric measurements for $\sim$ 1.7 billion sources.
Amongst them, 1.4 billion sources have \gb\ and \gr\ magnitudes and 1.3 billion sources 
have parallax and proper motion measurements. The $G$ band covers the 
whole optical wavelength range from 330 to 1050\,nm. The \gb\ and \gr\ magnitudes are derived from
the low resolution spectrophotometric measurements integrated over  
the wavelength ranges 330 - 680\,nm and 630 - 1050\,nm, respectively.
The internal validation of the Gaia DR2 shows that the calibration uncertainties for $G$, \gb\ and \gr\ are 
2, 5 and 3\,mmag, respectively.  

To break the degeneracy of effective temperature (or intrinsic colours) and extinction
for the individual stars, we combine the Gaia optical photometry 
with the IR photometry of 2MASS and WISE.
The 2MASS survey has three near-IR bands, $J,~H$ and \ks, 
centered at 1.25, 1.65 and 2.16\,$\mu$m, respectively. The 2MASS Point Source Catalog (PSC; 
\citealt{Skrutskie2006}) provides photometry for over 500 million objects. 
The uncertainties of 2MASS photometric measurements 
are estimated to be smaller than 0.03 mag. 
The WISE survey has four IR bands, $W1$ to $W4$, centered
at 3.4, 4.6, 12 and 22\,$\mu$m, respectively. The AllWISE Source Catalog \citep{Kirkpatrick2014} 
provides four band magnitudes and variability statistics for over 747 million objects.
For the WISE, we use only the data of band $W1$, as the longer wavelength measurements have 
lower sensitivities and poorer angular resolutions. Including the latter in the analysis  
do not improve the parameter estimation. 
The angular resolution of the WISE $W1$-band photometry is
 6.1$^{\prime\prime}$.

We select stars from the Gaia DR2 with Galactic latitude
$|b|$ $<$ 10\degr\ and then cross-match them with the 
2MASS PSC and the AllWISE Source catalogues
using the Centre de Donnes astronomiques de Strasbourg (CDS) XMatch 
Service\footnote{http://cdsxmatch.u-strasbg.fr/xmatch}.
The matching radius is set to 1.5$^{\prime\prime}$. 
The fraction of multiple matches is less than 0.01\,per\,cent and the matches of
the closest positions are adopted. We select sample stars by requiring that
the sources must be detected in all Gaia and 2MASS bands, i.e. Gaia $G$, \gb\ and \gr, and
2MASS $J$, $H$ and $K_{\rm S}$. We further require that the sources 
have photometric errors less than 0.08\,mag in all six bands and 2MASS quality of `AAA'.
The cuts lead to a total of 56,431,558 stars in the final catalogue.  
The Gaia $G$-band magnitude distribution of all the selected stars is illustrated in 
Fig.~\ref{datamd}. Most of the stars have a $G$-band magnitude between 8 and 18\,mag.
For a given star in the 
catalogue, its WISE $W1$-band magnitude will be adopted only when the 
photometric error is less than 0.08\,mag and no other stars are detected within 6.1$^{\prime\prime}$,
the angular resolution of the WISE $W1$ band in the combined photometric catalogue.
As a result, the $W1$-band magnitude is only used for just over 73\,per\,cent sources. The $G$-band
magnitude distribution of these latter stars is also plotted in Fig.~\ref{datamd}. 

\section{Method}

 \begin{figure*}
    \centering
  \includegraphics[width=0.99\textwidth]{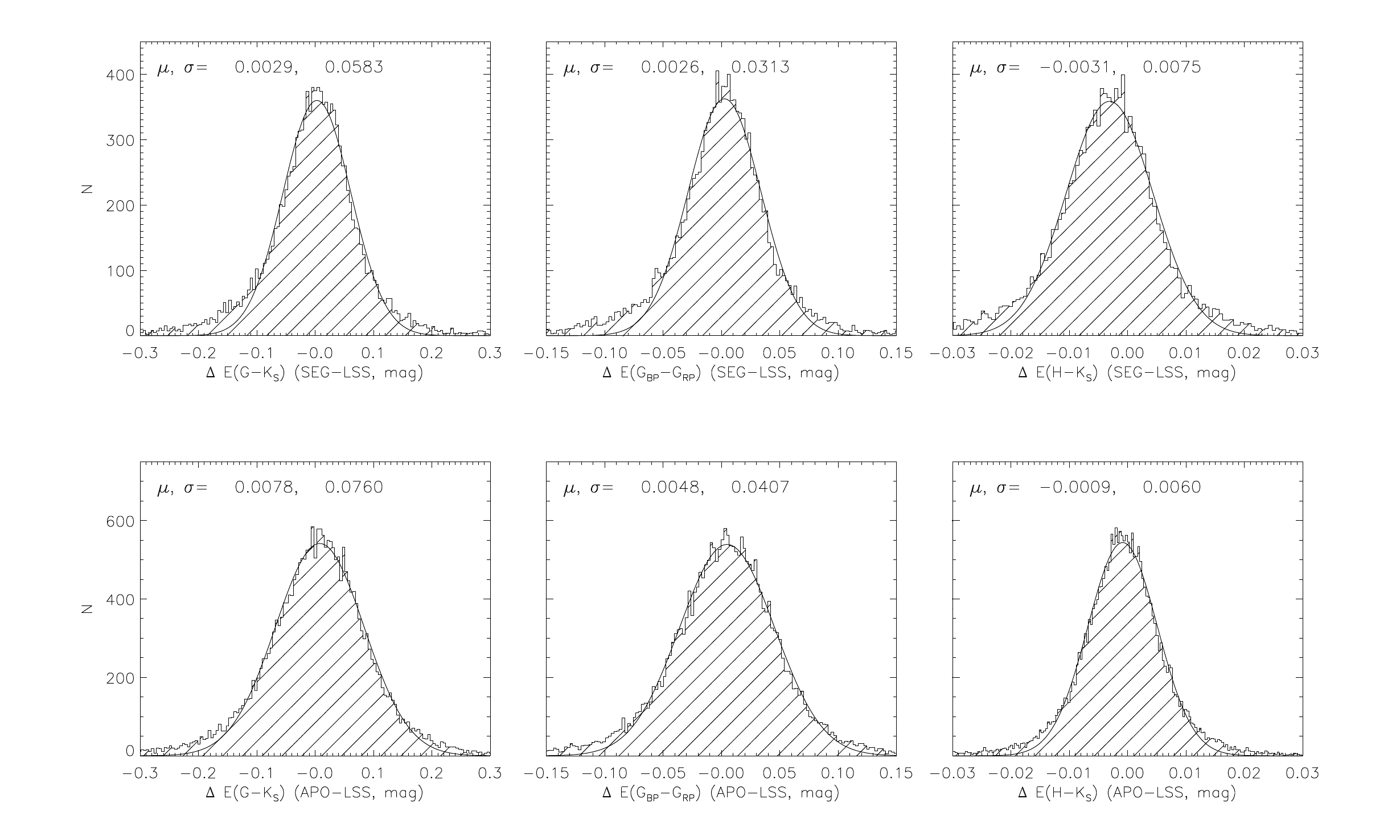}
  \caption{Differences of $E(G-K_{\rm S})$, \ebr\ and $E(H-K_{\rm S})$ values derived with the star-pair method
  for common stars between the LAMOST sample and the SEGUE (upper panels)
  and the APOGEE (bottom panels) samples.}
  \label{textc}
\end{figure*}

 \begin{figure}
    \centering
  \includegraphics[width=0.48\textwidth]{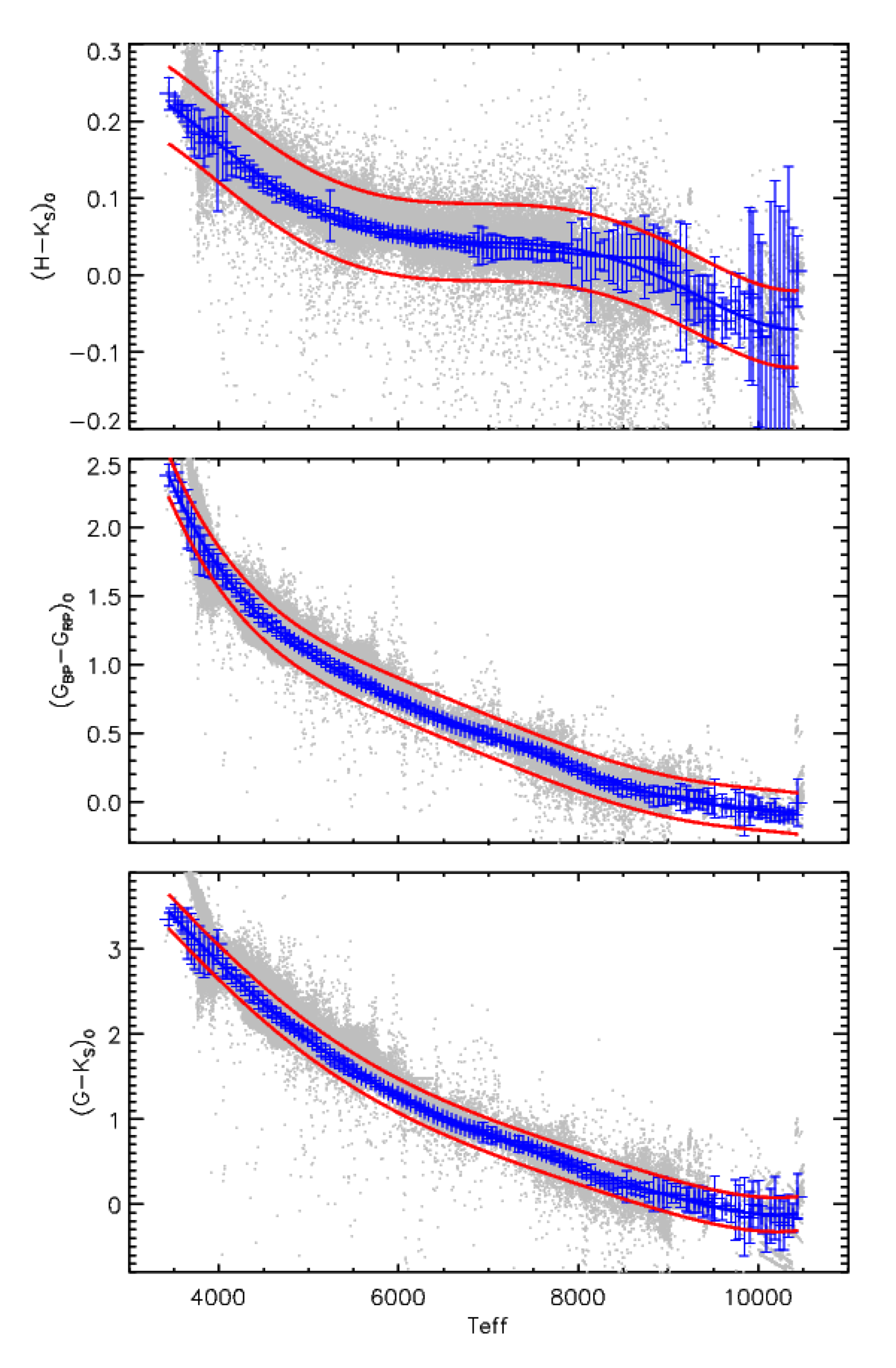}
  \caption{Intrinsic colours versus $T_{\rm eff}$ diagrams for stars in the LAMOST/SEGUE/APOGEE spectroscopic sample. 
Blue pluses and associated error bars represent median values and standard deviations deduced by binning the data points
in bins of 50\,K. The blue lines are fits to the median
values using a fifth-order polynomial.  
The red lines mark the boundaries used to reject the outliers that fall respectively more than 0.2, 0.15 and 0.05\,mag
from the fits.}
  \label{tte}
\end{figure}

 \begin{figure}
    \centering
  \includegraphics[width=0.48\textwidth]{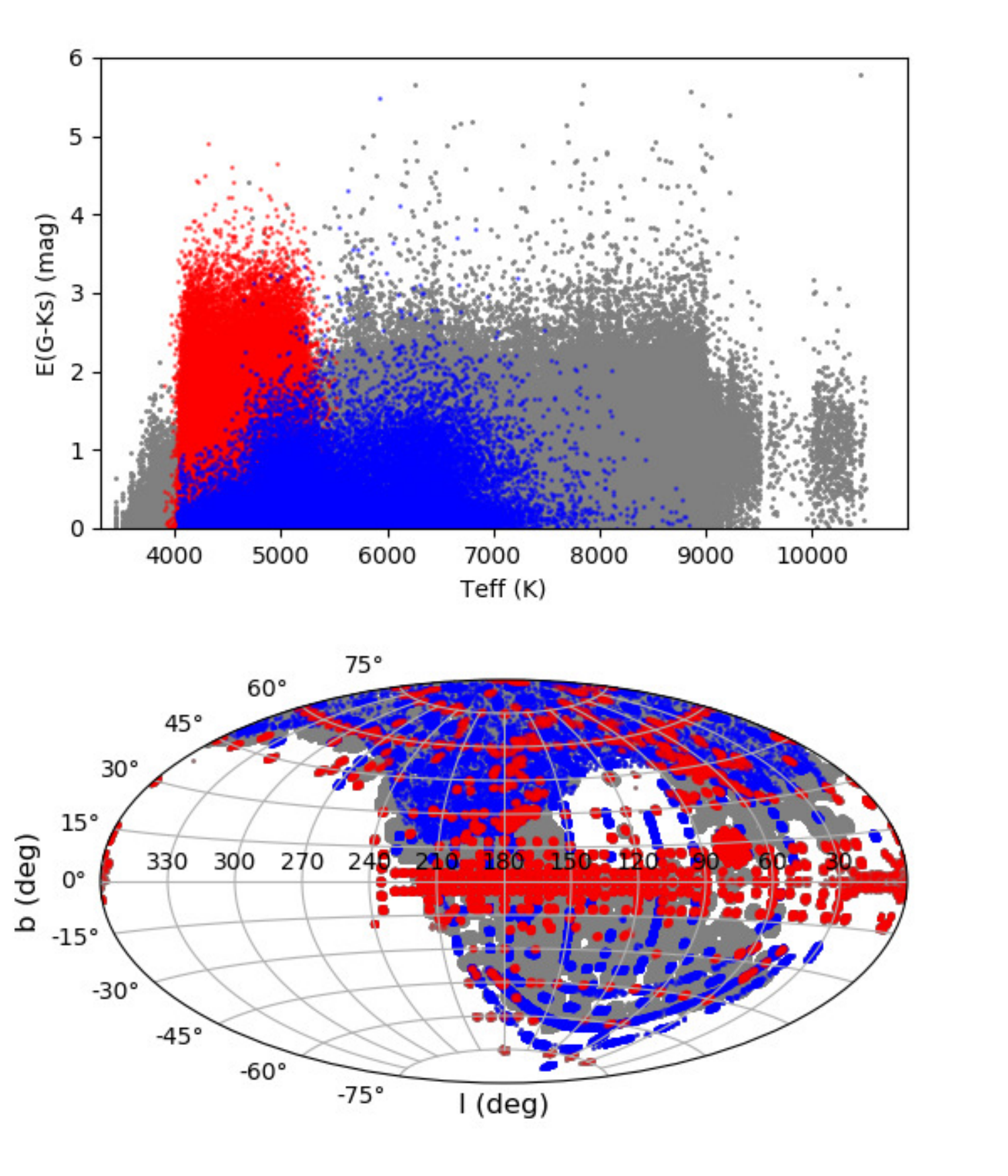}
  \caption{ Distributions of stars in the final spectroscopic sample in the Galactic coordinates (bottom panel) 
  and in the $E(G-K_{\rm S})$ versus $T_{\rm eff}$ plane (upper panel). 
  Grey, blue and red dots represent stars selected from 
 the LAMOST, SEGUE and APOGEE spectroscopic surveys, respectively.}
  \label{tspa}
\end{figure}

 \begin{figure*}
    \centering
  \includegraphics[width=0.98\textwidth]{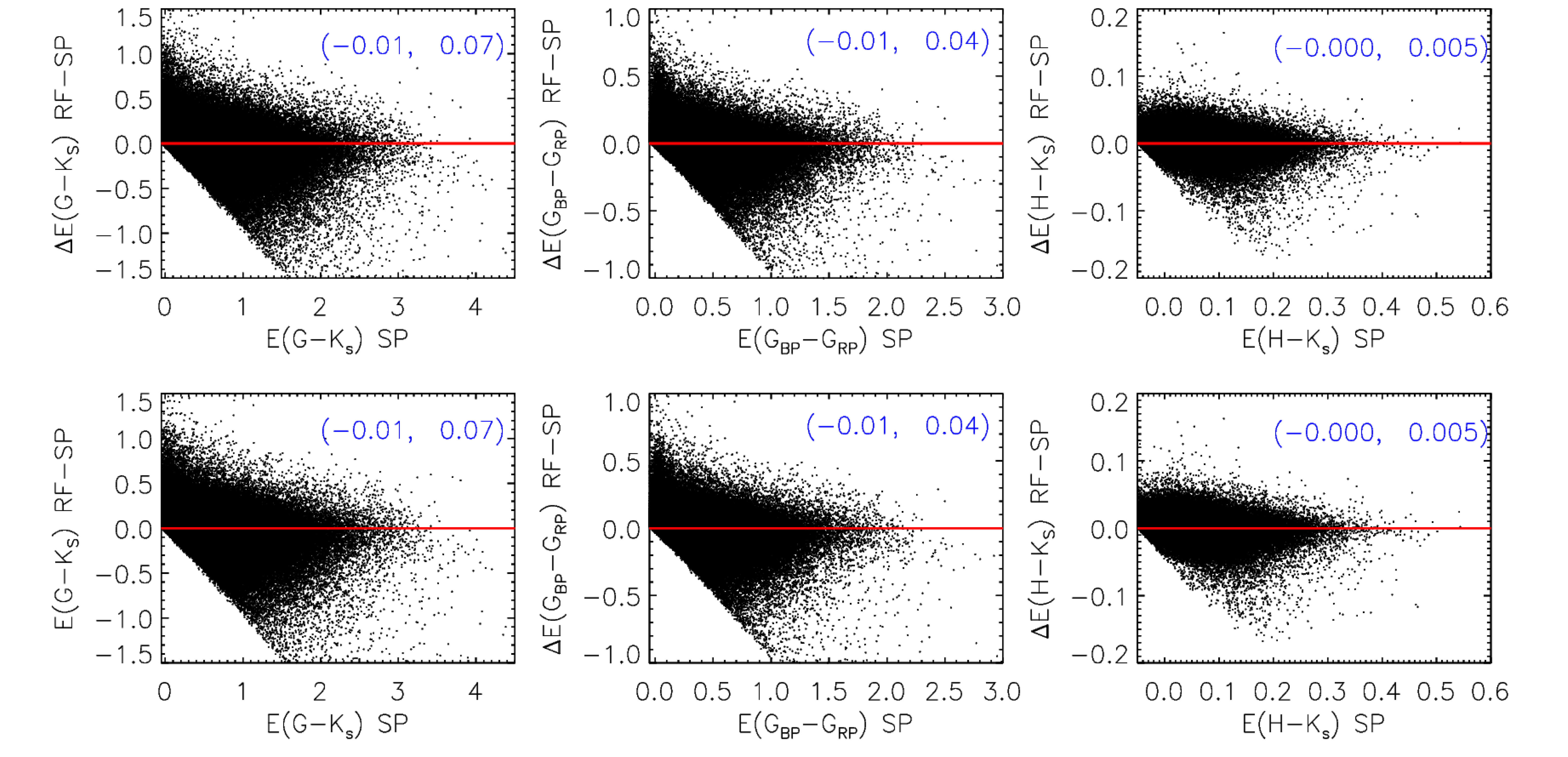}
  \caption{ Differences of colour excess values yielded by the Random Forest models and those derived with
 the star-pair technique for, from left to right,  $E(G-K_{\rm S})$, \ebr\ and $E(H-K_{\rm S})$,  respectively, for the
test sample stars. The upper and bottom three panels are those for the Random Forest models with and without
\ks $-$ $W1$ as input parameters, respectively.
The mean and standard deviation of the differences, are marked in each plot.}
  \label{rftest}
\end{figure*}

 \begin{figure*}
    \centering
  \includegraphics[width=0.98\textwidth]{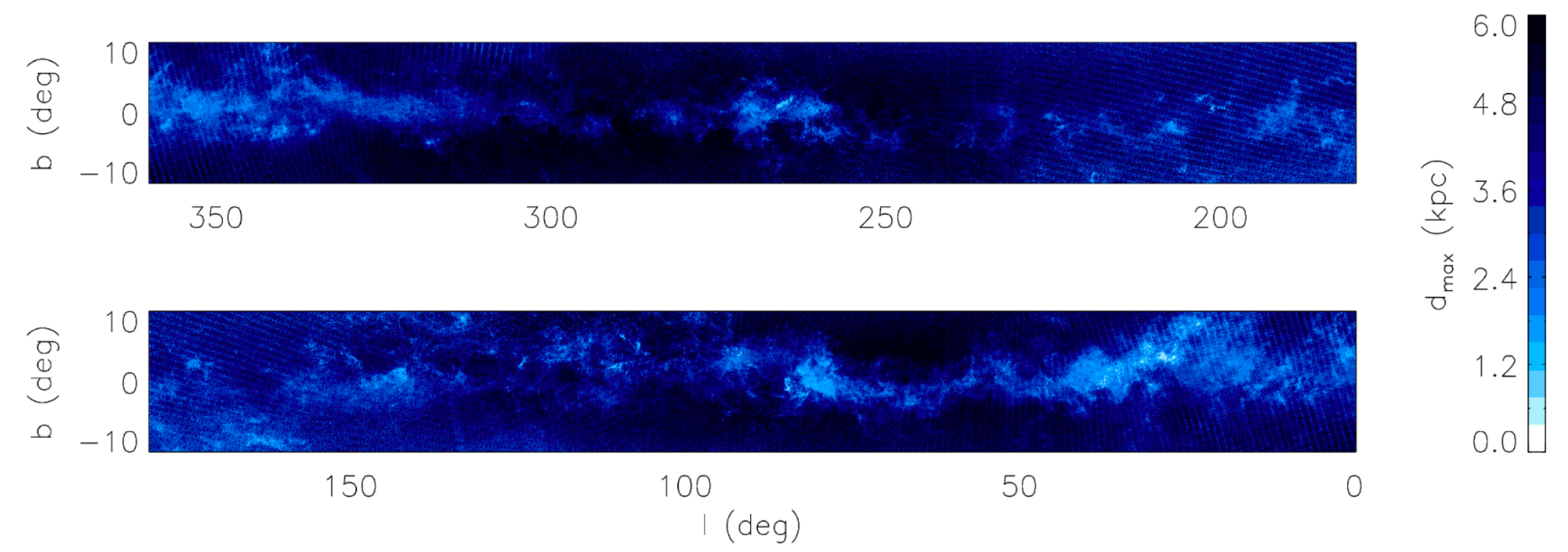}
  \caption{Map of the maximum distances for all pixels.}
  \label{dmax}
\end{figure*}

In the SED fitting or Bayesian approach, one uses gridding  
\citep{Berry2012, Chen2014} or the Markov chain Monte Carlo (MCMC) methods
\citep{Green2014, Sale2014} to sample the parameter space 
in order to derive the extinction (or colour excess) values of the individual stars. 
In the current work, we build a model that returns the colour excess values of stars
given the data described in Section 2. The model is based on a machine-learning algorithm,
the Random Forest regression, one of the most effective 
machine learning models for predictive analytics. 
A Random Forest regression is a meta estimator that fits a number of decision trees to various sub-samples of the 
dataset and uses average values to improve the predictive accuracy while controls over-fitting.  
We employ the Random Forest regression to estimate the colour excesses, 
\egk, \ebr\ and \ehk\, of the individual stars.
We have also tried other machine learning algorithms, 
such as the Support Vector Machine and Extra-Tree regressions 
 and found very similar results. 
 
 \subsection{ Training data set}

 We first create an empirical training set using stars from the spectroscopic surveys. 
Based on the stellar atmospheric parameters (T$_{\rm eff}$, log\,$g$ and [Fe/H]) from the spectra, we are able to
obtain accurate colour excesses of the individual stars.
We collect stars from the LAMOST and SDSS spectroscopic surveys.
A star-pair method is adopted to obtain the values of colour excesses of these stars.
The star-pair method is straightforward. It is based on the assumption that stars of
 the same stellar atmospheric parameters have 
the same intrinsic colours. Thus we can derive the intrinsic colours of a reddened star
from its pairs/counterparts of the same atmospheric parameters that suffer from either nil or well-known extinction.
In this work, we calculate values of \egk, \ebr\ and \ehk\ for stars from the spectroscopic surveys using the same 
star-pair algorithm of \citet{Yuan2015}. Firstly, a control sample\footnote{We note that this control sample is not the training data sample at all. The control sample is only used for the determination of colour excess of stars in the training sample.}
containing stars with nil or well-known extinction is built. We select stars of the control sample that 
suffer from very low values of extinction such that 
their $E(B-V)$ values can be well approximated by the reddening map of \citet[SFD]{Schlegel1998}.
For a given target star in the spectroscopic catalogues, its intrinsic colours, 
$(G-K_{\rm S})_0$, $(G_{\rm BP}-G_{\rm RP})_0$ and $(H-K_{\rm S})_0$, are estimated 
simultaneously from the corresponding values of the pair stars in the control sample. The values of 
\egk, \ebr\ and \ehk\ of the target star are then simply calculated from the differences between the observed and intrinsic 
colours, 
\begin{eqnarray}
&& E(G-K_{\rm S}) = (G-K_{\rm S}) - (G-K_{\rm S})_0,  \nonumber\\
&& E(G_{\rm BP}-G_{\rm RP}) = (G_{\rm BP}-G_{\rm RP}) - (G_{\rm BP}-G_{\rm RP})_0, \nonumber\\
&& E(H-K_{\rm S}) = (H-K_{\rm S}) - (H-K_{\rm S})_0. \nonumber
\end{eqnarray}  

The second release of value-added catalogues
of the LAMOST Spectroscopic Survey of the Galactic Anticentre (LSS-GAC DR2; \citealt{Xiang2017}) 
provides stellar atmospheric parameters deduced from 
1.8 million spectra. The catalogues for internal usage include all the observations collected by 
2016 June and contains robust 
stellar parameters estimated from over 5 million spectra \citep{Xiang2017, Chen2018}.
We select stars from the LSS-GAC DR2 with the criteria: 
LAMOST spectral S/N(4650\AA) per pixel  $>$ 10, 
`objtype' $=$ `STAR', `moondis'  $>$ 30\degr, `BADFIBER' $=$ 0, `SATFIBER' $=$ 0, `BRIGHTFIBER' $=$ 0,
`vr\_flag' $\le$ 6 and photometric errors 
err($G$, \gb, \gr, $J, ~H,$ \ks, $W1$) $<$ 0.1\,mag.
The requirements lead to 3,485,460 stars for the LAMOST sample.
Values of \egk, \ebr\ and \ehk\ are calculated for the LAMOST sample using the 
aforementioned star-pair technique.
The control sample for the LAMOST stars, which contains 48,243 stars,
is defined by requiring S/N(4650\AA) per pixel  $>$ 50, 
SFD $E(B-V)$ $<$ 0.015\,mag and err($G$, \gb, \gr, $J, ~H,$ \ks, $W1$) $<$ 0.03\,mag. 

We also use data from the SDSS DR14 that provides reliable stellar parameters for 0.4 million
stars from the Sloan Extension for Galactic Understanding and Exploration (SEGUE; \citealt{Yanny2009}) 
and 0.2 million stars from the  Apache Point Observatory Galactic Evolution Experiment 
(APOGEE; \citealt{Majewski2010}). Since the LAMOST surveys mainly the outer disc of the  
Galaxy, thus the SEGUE and APOGEE data sets complement that of LAMOST.
To minimize potential systematics between stellar parameters yielded by the different surveys, colour excesses 
\egk, \ebr\ and \ehk\, are calculated separately for the SEGUE and APOGEE samples, using the 
same aforementioned star-pair technique.
We select stars from the SEGUE and APOGEE catalogues with the criteria:  
S/N $>$ 10, $T_{\rm eff}$ $>$ 0, log\,$g$ $>$ 0, [Fe/H] or [M/H] $>$ $-3$ and 
err($G$, \gb, \gr, $J, ~H,$ \ks, $W1$) $<$ 0.1\,mag.
The requirements yield 116,006 and 142,994 stars for the SEGUE and APOGEE samples,
respectively. The SEGUE and APOGEE control samples are defined by requiring
S/N $>$ 50, SFD $E(B-V)$ $<$ 0.015\,mag and err($G$, \gb, \gr, $J, ~H,$ \ks, $W1$) $<$ 0.08\,mag.
The cuts yield 6,386 and 6,579 stars in the SEGUE and APOGEE control samples, respectively.

In Fig.~\ref{textc} we compare estimates of $E(G-K_{\rm S})$, \ebr\ and \ehk\ for common stars of the LAMOST and the 
SEGUE and APOGEE samples. We find no systematics amongst the different samples. For the LAMOST and SEGUE samples, 
the dispersions of differences are respectively 
$\sim$ 0.06, 0.03 and 0.008\,mag for $E(G-K_{\rm S})$, \ebr\ and \ehk\ differences.
The corresponding values are $\sim$ 0.08, 0.04 and 0.006\,mag between the LAMOST and APOGEE samples.
 
The resulted LAMOST, SEGUE and APOGEE samples are then combined. 
Abnormal colour excess estimates are excluded by plotting the intrinsic colours, 
$(G-K_{\rm S})_0$, $(G_{\rm BP}-G_{\rm RP})_0$ and $(H-K_{\rm S})_0$,
versus $T_{\rm eff}$ (Fig.~\ref{tte}), taking the advantage that the intrinsic colours of stars
correlates well with effective temperature. We reject stars that deviate more  
than 0.2, 0.15 and 0.05\,mag from the best-fit $(G-K_{\rm S})_0$, $(G_{\rm BP}-G_{\rm RP})_0$ and $(H-K_{\rm S})_0$ 
versus $T_{\rm eff}$ relations, respectively.  This leads us to a sample 
consisting 3,224,373 stars, which is adopted as the final training data set to
calculate the colour excesses of stars in the photometric sample by the Random Forest regression. 
The $G$-band magnitude distribution of the training sample is also plotted in Fig.~\ref{datamd}.
The magnitudes range between 8 and 19\,mag,  similar to the photometric sample.
In Fig.~\ref{tspa}, we show the  
distribution of these stars in the Galactic coordinates and
in the $E(G-K_{\rm S})$ versus $T_{\rm eff}$ plane. 
The sample covers almost all the Galactic latitudes and about two thirds of the Galactic longitudes. 
The effective temperatures range between 3000 and 10500\,K.  The estimated 
$E(G-K_{\rm S})$ values can be as high as $\sim$ 6\,mag, corresponding to $E(B-V)$ $\sim$ 3\,mag (based on the extinction
coefficients presented in Sect.~5.1). 
We note that the derived colour excesses for stars  
in the spectroscopic sample can be smaller than 0 but larger than
$\sim$ $-$0.05\,mag, as a result of the photometric errors.

\subsection{ Colour excesses determinations}

The stars in the final spectroscopic sample are randomly divided into two sub-samples, a training sample consisting of
80\,per\,cent of stars and a test sample containing the remaining 20\,per\,cent stars. 
The training sample is used to generate the Random
Forest models, while the test sample is used to validate the generated relations.
We build three separate Random Forest models for \egk, \ebr\ and \ehk, 
respectively. In all cases, the input parameters are
$G-J$, $G_{\rm BP}-K_{\rm S}$, $R_{BP}-J$, $J-H$ and $H-K_{\rm S}$. An extra colour $K_{\rm S} -W1$ will also 
be used for stars that have $W1$ data available.
We use the scikit-learn package for Python \citep{scikit-learn} to build the models.
We test with different parameters to optimize the models. Finally we 
set the number of trees in the forest to be n\_estimators=200, the 
number of features to consider when looking for the best split to be
max\_features=`auto', the minimum number of samples required to split an internal node 
to be min\_samples\_split=2 and the minimum number of samples needed at a leaf node 
to be min\_samples\_leaf=1. 
The 16th and 84th percentiles of the Random Forest ensemble are taken as the uncertainties. 
Fig.~\ref{rftest} compares the values of \egk, \ebr\ and \ehk\ yielded by the 
Random Forest models and those given by the
star-pair technique for the test sample stars. The Figure shows good agreement in all 
cases. There is no systematics in the residuals.  
The standard deviations of the differences are 0.07, 0.04 and 0.005 for \egk,
\ebr\ and \ehk, respectively, comparable to what expected from the typical uncertainties of 
colour excess of the spectroscopic sample.

 \subsection{3D colour excess distribution mapping}

For distances of the stars, we adopt the  
values from \citet{Bailer2018} who calculate distances of 1.3 billion stars from the  
Gaia measurements, imposing a prior based on the expected distribution of all stars in the Gaia catalogue. 
In the current work, we accept the distance estimates only for stars with Gaia parallax uncertainties 
smaller than 20\,per\,cent. This leads to a sample containing 35,354,103 stars. The $G$-band magnitude
distribution of these stars is also shown in Fig.~\ref{datamd}. Because of the cut in parallax errors, the number of 
stars drops dramatically at $\sim$18\,mag.

To map the 3D dust distribution in the Galactic plane, a procedure similar to that of \citet{Green2014} 
and \citet{Chen2017b} is adopted.
We divide the sample stars into subfields (pixels) of size 6 $\times$ 6\,arcmin.
For each pixel, we define the reliable 
depth of our extinction map as the maximum distance of all the
individual stars in that pixel. In Fig.~\ref{dmax} we present a map of those depths. The maximum distance
is typically 4\,kpc, rising up to 6\,kpc for regions of the smallest dust extinction. There
are some stripe-like features in the map. They are artifacts caused by the scanning nature 
of the Gaia observations.

 \begin{figure}
    \centering
  \includegraphics[width=0.48\textwidth]{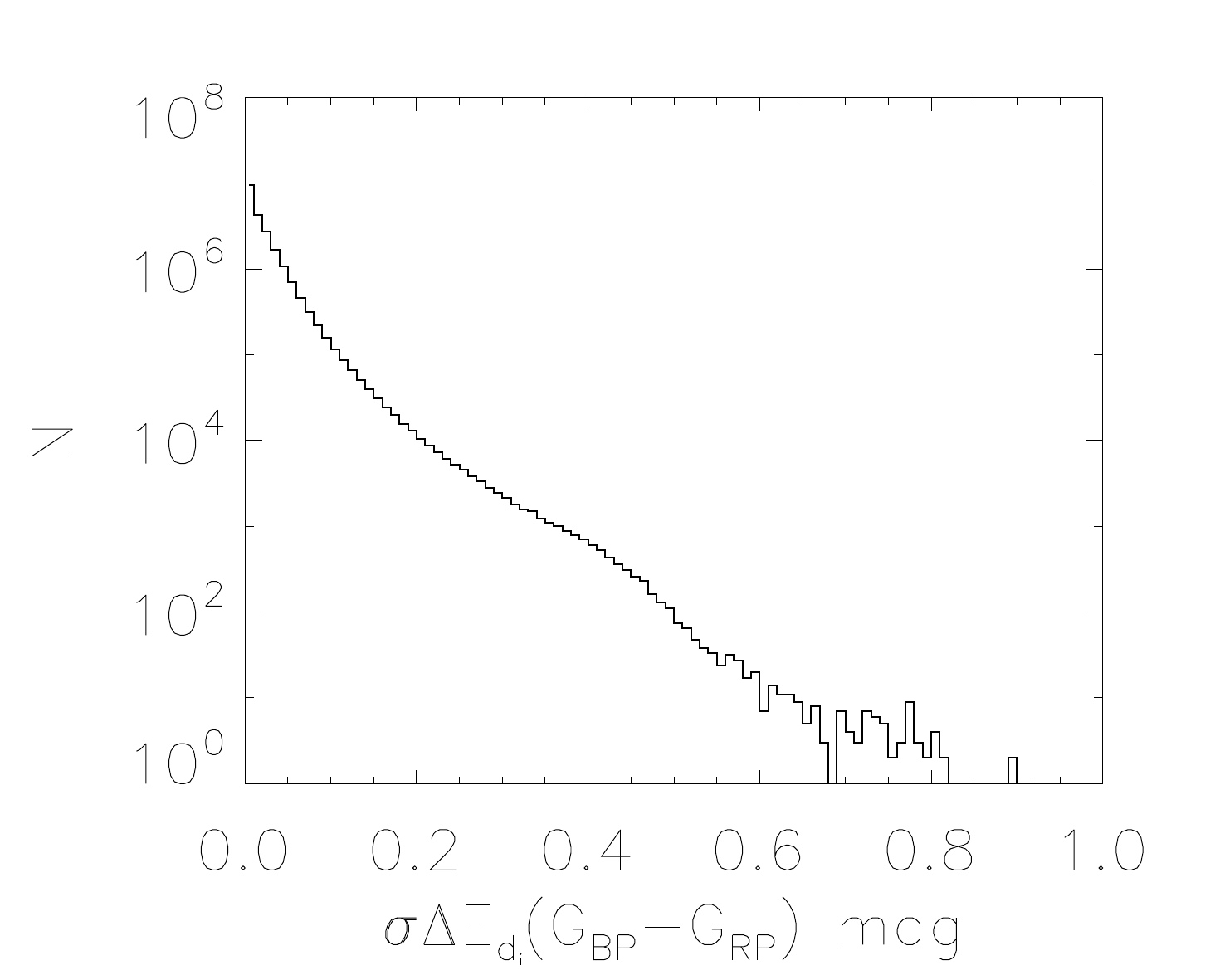}
  \caption{Histogram distribution of the widths of the 68\,per\,cent confidence intervals of 
  $\Delta E_{d_i}(G_{\rm BP}-G_{\rm RP})$.}
  \label{exterr}
\end{figure}

For each pixel, the colour excess profile $E(d)$ as a function of distance $d$ is then derived by 
fitting the colour excess as a function of distance
for the individual stars in that pixel. Here $E$ refers to colour excess 
\egk, \ebr\ or \ehk.  At a given distance $d$, the colour excess $E(d)$ is parameterized by a piecewise linear function, 
\begin{equation}
E(d) = \Sigma_{d_i=0}^{d_i=d} (\Delta E_{d_i}),
\end{equation} 
where $\Delta  E_{d_i}$ is colour excess produced by the local dust grains in the $i$-th distance bin ($d_i$
is the distance of the bin).
$\Delta E_{d_i}$ is set to be no less than zero to prevent negative reddenings.
The length of each distance bin is set as $\delta d = 0.2$\,kpc. 

Assuming a set of $\Delta E_{d_i}$ for a given pixel,
we are able to model the colour excesses $E^n_{\rm mod}$ of any star of index $n$ in the pixel from Eq.~(1).
We perform an MCMC analysis to 
find the best set of $\Delta E_{d_i}$ of the pixel that maximise the likelihood defined as,
\begin{equation}
L = \Pi_{n=1}^{N} \frac{1}{\sqrt{2\pi} \sigma_n}{\exp}(\frac{-(E^n_{\rm obs}-E^n_{\rm mod})^2}{2\sigma^2_n}),
\end{equation}
where $n$ is index of star in the pixel, 
$E^n_{\rm obs}$ and $E^n_{\rm mod}$ are respectively the colour excess derived in
Sect.~3.2 and that given by Eq.~(1) of the star, $\sigma_n$ is the combined
uncertainty of the derived colour excess ($\sigma_{E_{\rm obs}^n}$) and distance ($\sigma_{d^n}$), 
given by $\sigma_n = \sqrt{\sigma_{E_{\rm obs}^n}^2+(E\dfrac{\sigma_{d^n}}{d^n})^2}$, 
$d^n$ is the distance of the star \citep{Lallement2014, Chen2017b}), 
and $N$ is the total number of stars in the pixel. We note that the error 
$E\dfrac{\sigma_{d^n}}{d^n}$  resulting
from the distance error is only an approximation under the
assumption that the dust opacity is constant along the line of sight.
In the current work, the
distance uncertainties are simply adopted as $\sigma _d = \dfrac{d_{\rm hi} - d_{\rm lo}}{2}$, where $d_{\rm hi}$ 
and $d_{\rm lo}$ are respectively the upper and lower bounds of the 68\,per\,cent confidence interval
of the distance estimate from \citet{Bailer2018}.
The median number of stars in a pixel is 45. 
About 1000 ($\sim$ 0.1\,pre\,cent) pixels do not have enough stars ($N$ $<$ 3) to 
obtain the colour excess profile. We have enlarged those pixels to size of 15 $\times$ 15\,arcmin.
The uncertainties of  $\Delta E_{d_i}$
are computed from 68\,per\,cent probability intervals of
the marginalized probability distribution functions (PDFs) of each
parameter, given by the accepted values after post-burn period in
the MCMC chain. Histogram distribution
of the resulted uncertainties of $\Delta E_{d_i}(G_{\rm BP}-G_{\rm RP})$ 
is shown in Fig.~\ref{exterr}.\\

\section{Dust reddening maps}

\begin{table*}
   \centering
  \caption{Description of the 3D dust reddening data of the Galactic disk.}
  \begin{tabular}{ccl}
  \hline
  \hline
Column & Name & Description  \\
\hline
1 & $l$ & Galactic longitude (\degr) of the pixel  \\
2 & $b$ & Galactic latitude (\degr) of the pixel  \\
3 & $d_{\rm max}$ & Reliable depth (kpc) of the pixel  \\
4 $-$33 &   \egk$_{\rm 0.2-6\,kpc}$ & Integrated \egk\ as a function of distance $d_i$ ranging from 0.2 to 6\,kpc with a step of 0.2\,kpc \\
34 $-$63 &   $\sigma$ \egk$_{\rm 0.2-6\,kpc}$ &  Errors of \egk\ \\
64 $-$93 &   \ebr$_{\rm 0.2-6\,kpc}$ & Integrated \ebr\ as a function of distance $d_i$ ranging from 0.2 to 6\,kpc with a step of 0.2\,kpc   \\ 
94 $-$123 &   $\sigma$\ebr$_{\rm 0.2-6\,kpc}$ &  Errors of \ebr\ \\
124 $-$153 &   \ehk$_{\rm 0.2-6\,kpc}$ &Integrated \ehk as a function of distance $d_i$ ranging from 0.2 to 6\,kpc with a step of 0.2\,kpc  \\ 
154 $-$183 &   $\sigma$\ehk$_{\rm 0.2-6\,kpc}$ & Errors of \ehk\    \\
\hline
\end{tabular} 
\end{table*}

 \begin{figure}
    \centering
  \includegraphics[width=0.47\textwidth]{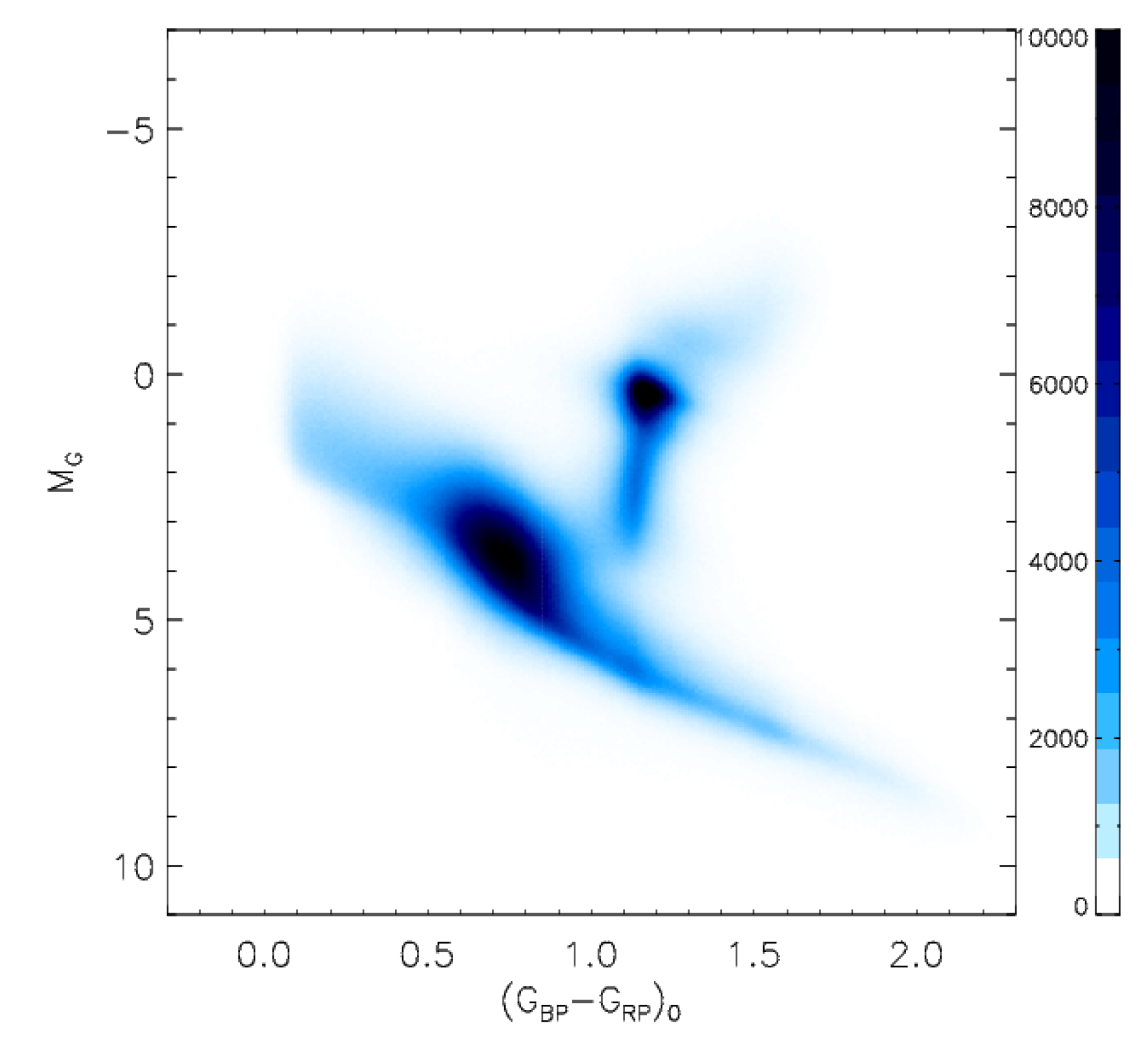}
  \caption{Dust corrected colour and absolute magnitude diagram of all stars with Gaia parallax errors smaller than 
  20\,per\,cent in the combined  photometric sample. The colour scale 
  represents the number of stars per colour-magnitude bin of size 0.01 $\times$ 0.05\,mag.}
  \label{cmdp}
\end{figure}

 \begin{figure*}
    \centering
  \includegraphics[width=0.95\textwidth]{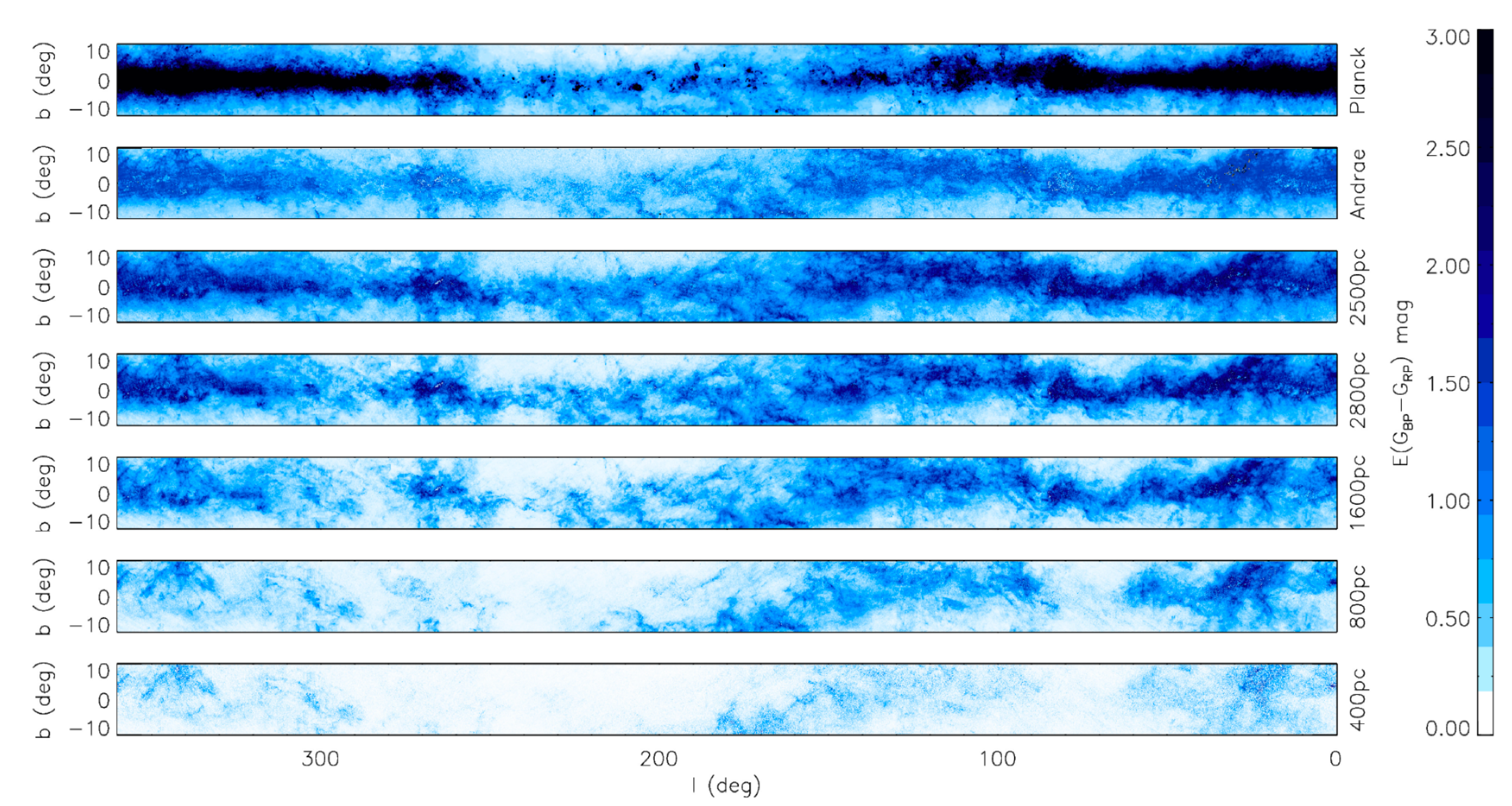}
  \caption{Slices of 2D cumulative colour excess maps of the Galactic plane, integrated from the 3D maps to  
 distances, from bottom to the third from the top, 400, 800, 1600, 2800 and 5000\,pc. 
 The second panel from the top shows the distribution of median \ebr\ from \citet{Andrae2018} 
 and the top panel the 2D map from \citet{Planck2014} for comparison. For the 
Planck map, the \ebr\ values are converted from
values of $E(B-V)$ using the extinction coefficients given in Section~5.1 of the current work. Colour excess values
 larger than 3\,mag are represented by black in the diagrams.}
  \label{extint}
\end{figure*}

 \begin{figure*}
    \centering
  \includegraphics[width=0.98\textwidth]{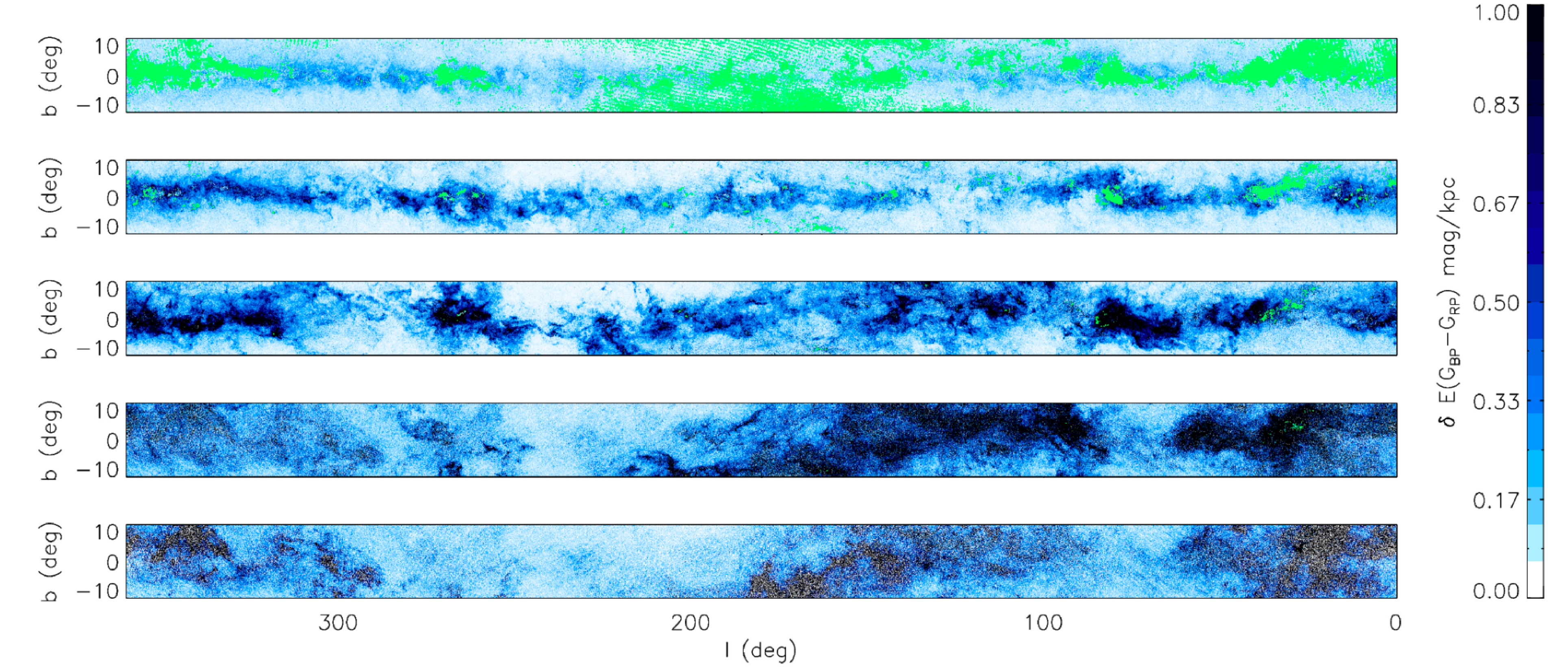}
  \caption{Distributions of differential colour excesses  $\delta$\ebr\ in the Galactic plane, in
units of mag\,kpc$^{-1}$.  The five panels, from bottom to top, refer to ranges of distance from the Sun, 
0 - 400\,pc, 400 - 800\,pc, 800 - 1600\,pc, 1600 - 2800\,pc and 2800 - 5000\,pc, respectively.  
The differential colour excess values larger than 1\,mag\,kpc$^{-1}$ are plotted in black in the diagrams.
The green regions are those beyond the maximum 
reliable distances pixels.}
  \label{extsli}
\end{figure*}

 \begin{figure*}
    \centering
  \includegraphics[width=0.98\textwidth]{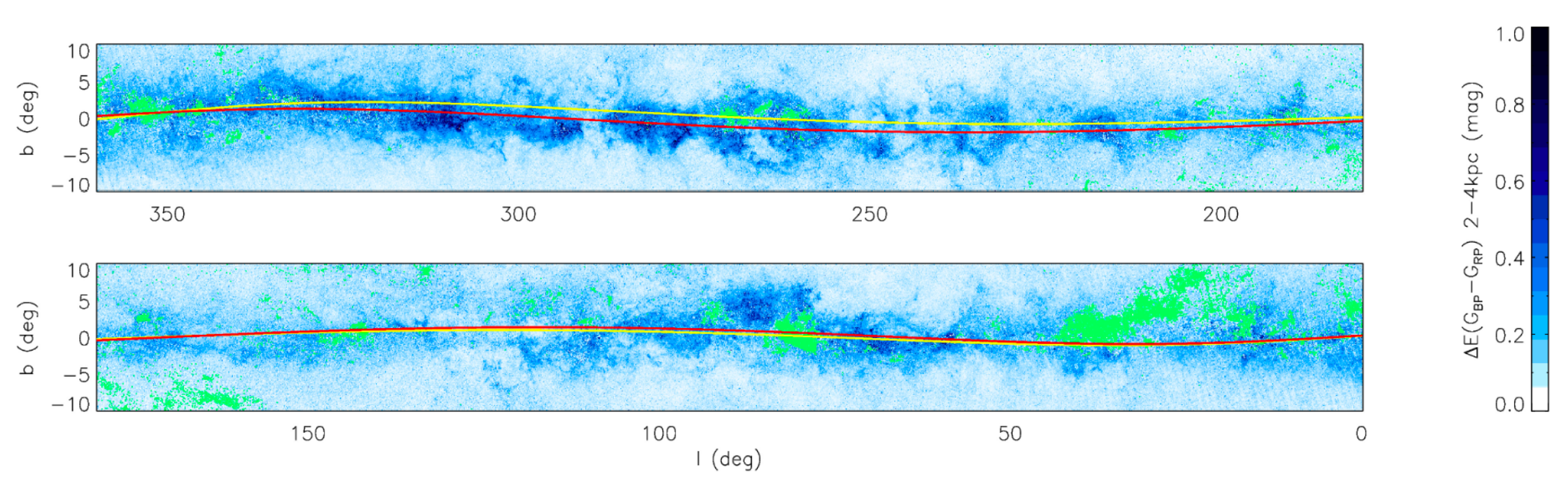}
  \caption{Warp in the dust distribution as revealed by our 3D colour excess map. The plot shows  
  the differential colour excess $\Delta$\ebr\ in the Galactic plane produced by dust in distance bin from 2 to 4\,kpc from the Sun. 
  The green regions are those beyond the maximum reliable distances pixels.
 The red line delineates the best fit to our data. The yellow line represents the result of \citet{Marshall2006}.}
  \label{warp}
\end{figure*}

 \begin{figure*}
    \centering
  \includegraphics[width=0.98\textwidth]{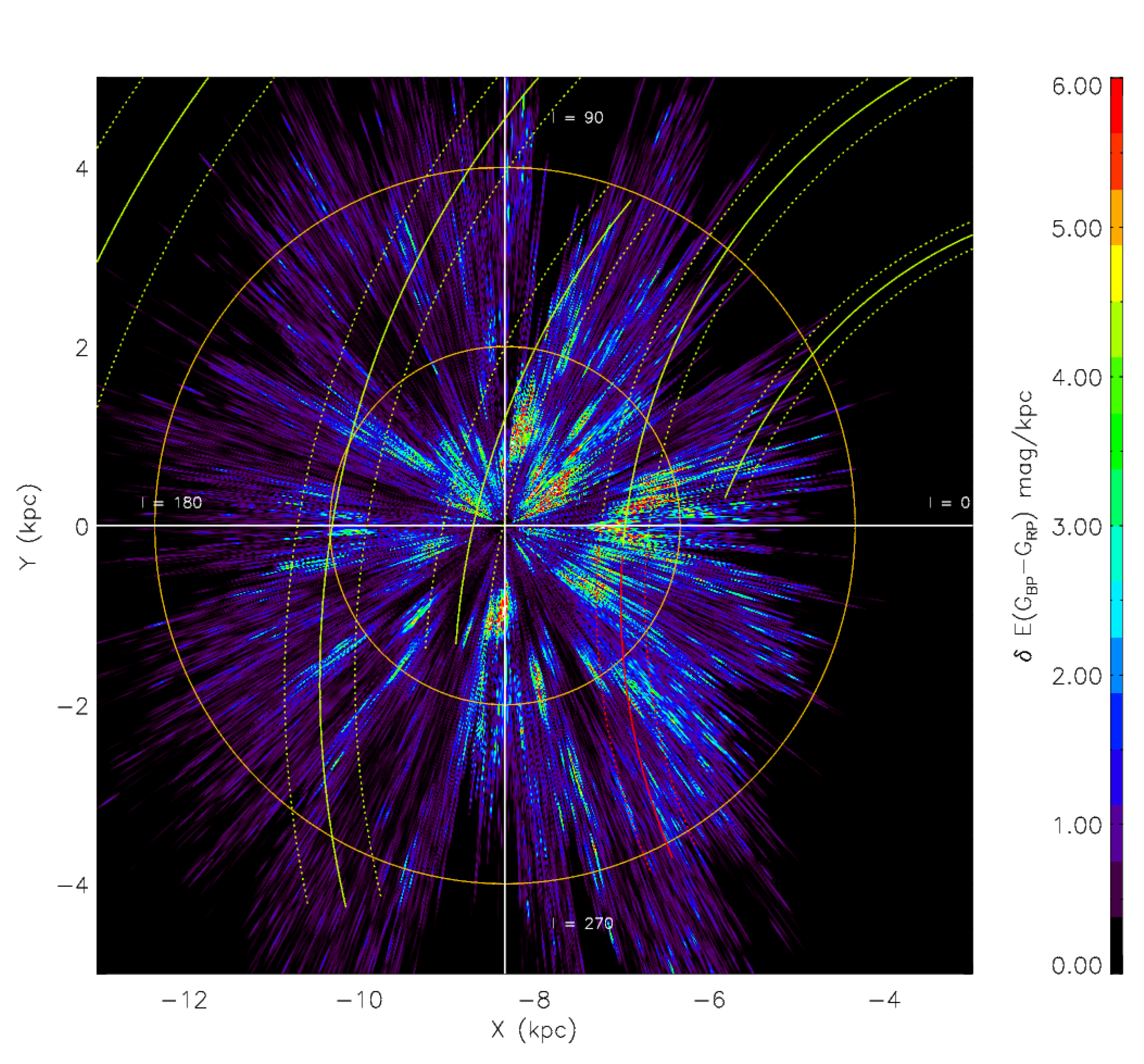}
  \caption{Distribution of local dust in the Galactic plane ($|b|$ $<$ 0.1\degr). 
The Sun, assumed to be at 8.34\,kpc from the Galactic center, is located at the centre of the plot. 
Brown circles are placed every 2 kpc from the Sun.  The directions of $l$ = 0\degr, 90\degr, 180\degr\ and 
270\degr\ are also marked in the plot. The yellow solid and dashed curved lines denote the center and 
$\pm$1\,$\sigma$ widths of spiral arm models of, from left to right, the Outer, Perseus, Local,  Sagittarius and Scutum arms, 
taken from Reid et al. (2014).  The red lines are the same model of Sagittarius but have Galactocentric azimuth $\beta$ 
extending from $-$2\degr\ to $-$30\degr. 
}
  \label{extb0}
\end{figure*}

 Our resulted 3D colour excess maps are available
in electronic form in the online version of this manuscript\footnote{Table~1 is also available online via \\
``http://paperdata.china-vo.org/diskec/cestar/table1.zip''.}.
Table~1 describes the data format. Each row of the catalogue contains
information of one subfield (pixel): Galactic coordinates $l$ and $b$, the reliable depth $d_{\rm max}$, 
measured values of the integrated \egk, \ebr\ and \ehk\ and the associated
uncertainties at the individual distance bins. In addition, the colour excesses, \egk, \ebr\ and \ehk\ and
the associated errors of each star in the combined photometric sample are
available upon request by email. 
Fig.~\ref{cmdp} plots the dust corrected colour and absolute magnitude diagram for 
over 35 million stars with Gaia parallax errors smaller than 20\,per\,cent.
Values of absolute magnitude $M_G$ of the individual stars are estimated using the
standard relation, $M_G = G-A_G-5{\rm log}d+5$, where $d$ is distance estimate from
\citet{Bailer2018} and $A_G$ the $G$-band line-of-sight extinction.  As the Gaia $G$ band covers
a wide wavelength range, we calculate $A_G$ from colour excess
\egk\ and line of sight extinction $A_{K_{\rm s}}$, as $A_G = E(G-K_{\rm S})+A_{K_{\rm S}}$, where
$A_{K_{\rm s}}$ is obtained from
\ehk\ using the near-IR extinction law of \citet{Yuan2013}, $A_{K_{\rm S}} = 1.987E(H-K_{\rm S})$. So we have 
\begin{equation}
A_G = E(G-K_{\rm S})+1.987E(H-K_{\rm S}).	
\end{equation}
Fig.~\ref{cmdp} presents a nice Hertzsprung-Russell diagram (HRD) very similar to 
the HRD presented in Fig.~5 of \citet{Gaia2018b} which is constructed using low-extinction stars.
The main sequence is quite sharp, the red clump really a clump and the red giant branch 
clearly visible. The well-defined HRD presented in Fig.~\ref{cmdp} suggests the 
robustness of colour excess values derived in the current work.

We plot in Fig.~\ref{extint} 2D maps of colour excess \ebr\ in the Galactic plane,
integrated respectively to selected distances, 400, 800, 1600, 2800 and 5000\,pc from the Sun. 
In general, the colour excess increases with distance for
all pixels, but the growth rate varies from pixel to pixel, showing
various structures.  At close distances, we see the local dust clouds that 
extend to the high latitudes. At large distance, we begin to see the tilt of the dust lane in the Galactic disk.
Also plotted in the Figure for comparisons are the distribution of median \ebr\ values from \citet{Andrae2018}
and the Planck 2D colour excess map
deduced from the dust far-IR thermal emission \citep{Planck2014}.
We obtain the median \ebr\ values for each pixel based on the estimates from \citet{Andrae2018}. 
There is high degree of similarity of the overall structure and features between
the Andrae et al. median \ebr\  map and our maps.
The Planck map, representing the colour excess integrated along the line-of-sight to infinite, is comparable with 
ours integrated to 4\,kpc in the direction of outer disk (150\degr\ $<$ $l$ $<$ 250\degr). 
But in the direction toward the Galaxy centre, the Planck
map yields systematically much larger colour excess values than ours, suggesting that there are 
still large amounts of dust in that direction beyond 4\,kpc. Nevertheless, both maps show 
very similar dust features.
 
To highlight dust features in different distance slices, we plot in
Fig.~\ref{extsli} the distribution  of the differential colour excess $\delta$\ebr\ (in units of mag\,kpc$^{-1}$) 
produced by the local dust grains in distance slices 0 $-$ 400\,pc,  
400 $-$ 800\,pc, 800 $-$ 1600\,pc, 1600 $-$ 2800\,pc and 2800 $-$ 5000\,pc, respectively.
Only the map of $\delta$\ebr\ is shown, as the maps of $\delta$\egk\ and $\delta$\ehk\ are very similar.
  At large distances, the dust reddening in the disk is smaller than that at smaller distances. 
This is mainly due to the selection bias of the map.
Distant stars suffering from larger dust extinction are fainter and have larger photometric/distance uncertainties 
compared to those suffering from smaller dust extinction. 
Those highly reddened distant stars are therefore discriminated against 
when constructing the 3D colour excess maps. 
Fig.~\ref{extsli} shows fine structures of dust distribution at various distances bins.
The local dust clouds are clearly visible in the two closest distance slices ($d$ $<$ 800\,pc). For example, 
in these two slices, one sees the Ophiuchus ($l$ $\sim$ 345\degr\ $-$ 10\degr), 
Aquila Rift ($l$ $\sim$ 20\degr\ $-$ 40\degr), and Hercules  ($l$ $\sim$ 40\degr\ $-$ 50\degr) in the
inner disk, Polaris Flare ($l$ $\sim$ 120\degr\ $-$ 130\degr), Cepheus Flare  ($l$ $\sim$ 100\degr\ $-$ 115\degr)
 and Perseus-Taurus-Auriga complex  ($l$ $\sim$ 150\degr\ $-$ 185\degr) in the direction
of the Galactic anti-centre and the Gum Nebula at $l$ $\sim$ 260\degr.
Beyond 800\,pc, one sees mainly the dust features in the Galactic thin disc, concentrated in  
a narrow range of latitude ($|b|$ $<$ 5\degr), and the warp is clearly
visible. Following \citet{Marshall2006}, we fit the dust warp by the equation,
\begin{equation}
z_{\rm warp} = \gamma (R - R_0) \cos (\theta - \theta_0),
\end{equation}
where $z_{\rm warp}$ is the vertical distance between the mid-plane of the dust disk and the plane
defined by $b$ = 0\degr, $\gamma$ the slope
of the amplitude, $R_0$ the Galactocentric radius where the warp starts and $\theta_0$
the node angle. Based on the distribution of dust grains between distances from 2 to 4\,kpc (Fig.~\ref{warp}), 
we find the values of position $z$ of maximum colour excess in the individual ($R$, $\theta$) bins. 
Parameters $\gamma$, $R_0$ and $\theta_0$ are then derived by fitting the results. The best-fit values are
$\gamma$ = 0.13, $R_0$ = 7.3\,kpc and $\theta_0$ = 93\degr. The fit is shown in 
Fig.~\ref{warp}, with the result of \citet{Marshall2006} over-plotted for comparison. Our
fit is in good agreement with that of \citet{Marshall2006} in the common region that is also covered by the 
dust map of Marshall et al. ($-$100\degr\ $<$ $l$ $<$ 100\degr). However, in other regions such as 
170\degr\ $<$ $l$ $<$ 250\degr, there is significant deviation between the warp model of Marshall et al. 
and ours. 

In Fig.~\ref{extb0} we show the differential colour-excess $\delta$\ebr\ (in units of 
mag\,kpc$^{-1}$) in the Galactic disc plane of
$|b|$ $<$  0.1\degr. The Sun is located at the center of the plot at $X$ = $-$8.34\,kpc and $Y$ = 0\,kpc.
The overall morphology of the dust structure inside 2\,kpc matches well with 
those of Fig.~1 in \citet{Lallement2014} and of Fig.~7 in \citet{Green2018}.
One can easily locate the various  
dust clouds, such as the  Aquila Rift complex and the Cygnus Rift in the Figure.
On large scales, those clouds are likely to be spatially coincident with the Galactic 
spiral arms delineated by log-spiral fits to the high-mass star forming regions of \citet{Reid2014}. 
The Sagittarius, Local and Perseus arms are discernible in our map. 
The Scutum and Outer arms locate respectively near the right and left edges of the map 
and are not readily identifiable. 
The Perseus arm seems to be traced by several discrete dust clouds 
located at $\sim$2\,kpc from the Sun in the outer disk. The Sagittarius arm is probably traced by 
several clouds located at $R$ about 6 - 7\,kpc. 
In addition, the Sagittarius arm seems to be quite
extended in the fourth quadrant as traced by a few clouds in the directions
between $l$ $\sim$ 310\degr\ and 360\degr, 
consistent with the most recently work of  \citet{Xu2018} who study the spiral arm structure in the solar neighborhood 
using a sample of over 5000 O-B2 stars. The Local arm is likely associated with the 
several discrete clouds at $R$ between 7 and 9\,kpc. 

\section{Discussion}

\subsection{Extinction coefficients for the Gaia photometry}

 \begin{figure}
    \centering
  \includegraphics[width=0.48\textwidth]{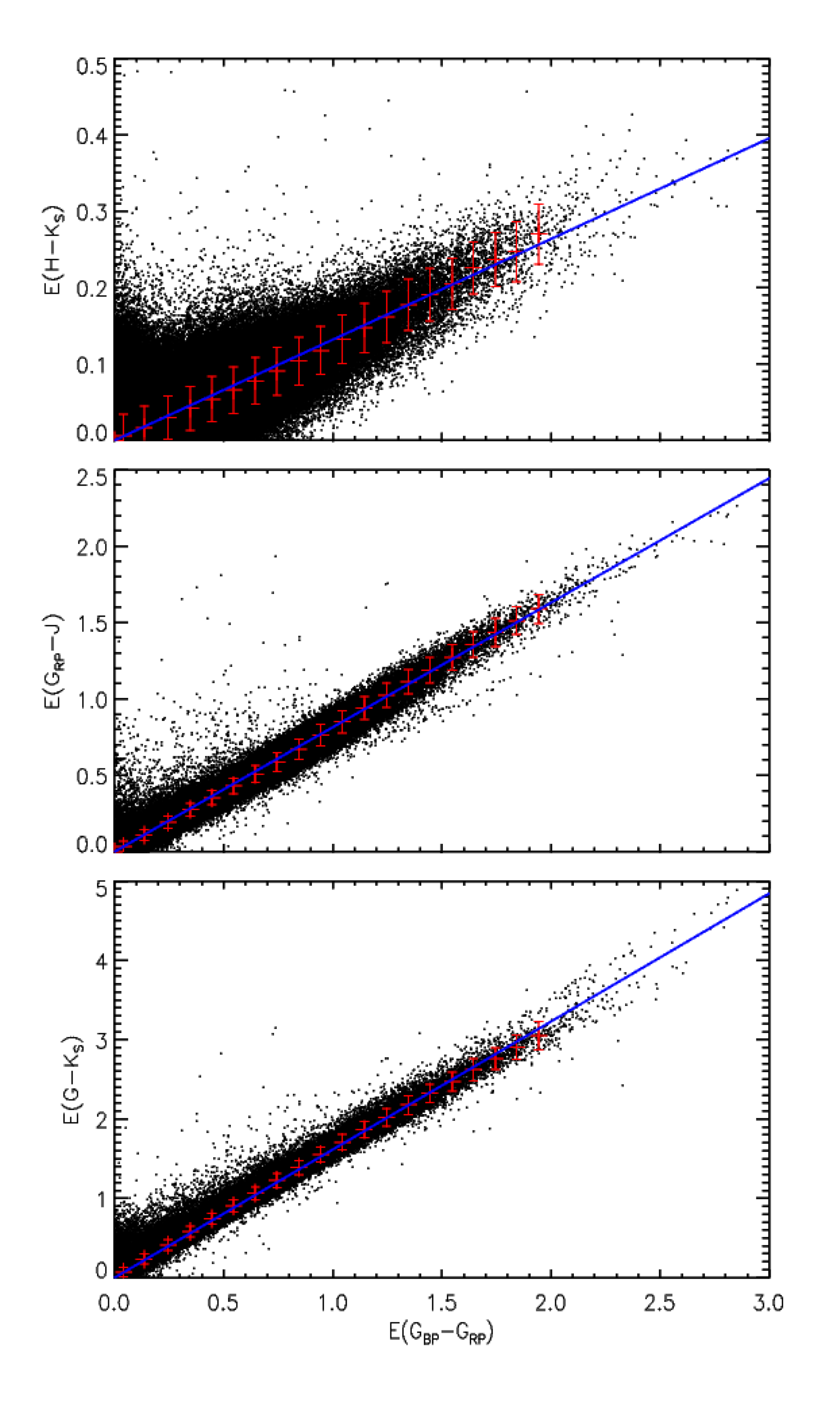}
  \caption{Colour excesses $E(G-K_{\rm S})$, $E(G_{\rm RP}-J)$ and $E(H-K_{\rm S})$ versus $E(G_{\rm BP}-G_{\rm RP})$. 
Red pluses and error bars represent medians and dispersions in bins of size 0.1\,mag in abscissa. 
Blue lines are linear regressions passing through the origin to the red pluses.}
  \label{extl}
\end{figure}

\begin{table*}
   \centering
  \caption{Colour excess ratios and extinction coefficients for Gaia photometry}
  \begin{tabular}{cccccc}
  \hline
  \hline
$\dfrac{E(G-K_{\rm S})}{E(G_{\rm BP}-G_{\rm RP})}$ & $\dfrac{E(G_{\rm RP}-J)}{E(G_{\rm BP}-G_{\rm RP})}$ & $\dfrac{E(H-K_{\rm S})}{E(G_{\rm BP}-G_{\rm RP})}$ & $R_G$ & $R_{G_{\rm BP}}$ & $R_{G_{\rm RP}}$ \\
\hline
1.61 $\pm$ 0.02 & 0.81 $\pm$ 0.02 & 0.13 $\pm$ 0.01 & 2.50 $\pm$ 0.03 &3.24 $\pm$ 0.02 & 1.91 $\pm$ 0.02 \\
\hline
\end{tabular}
\end{table*}

 \begin{figure*}
    \centering
  \includegraphics[width=0.98\textwidth]{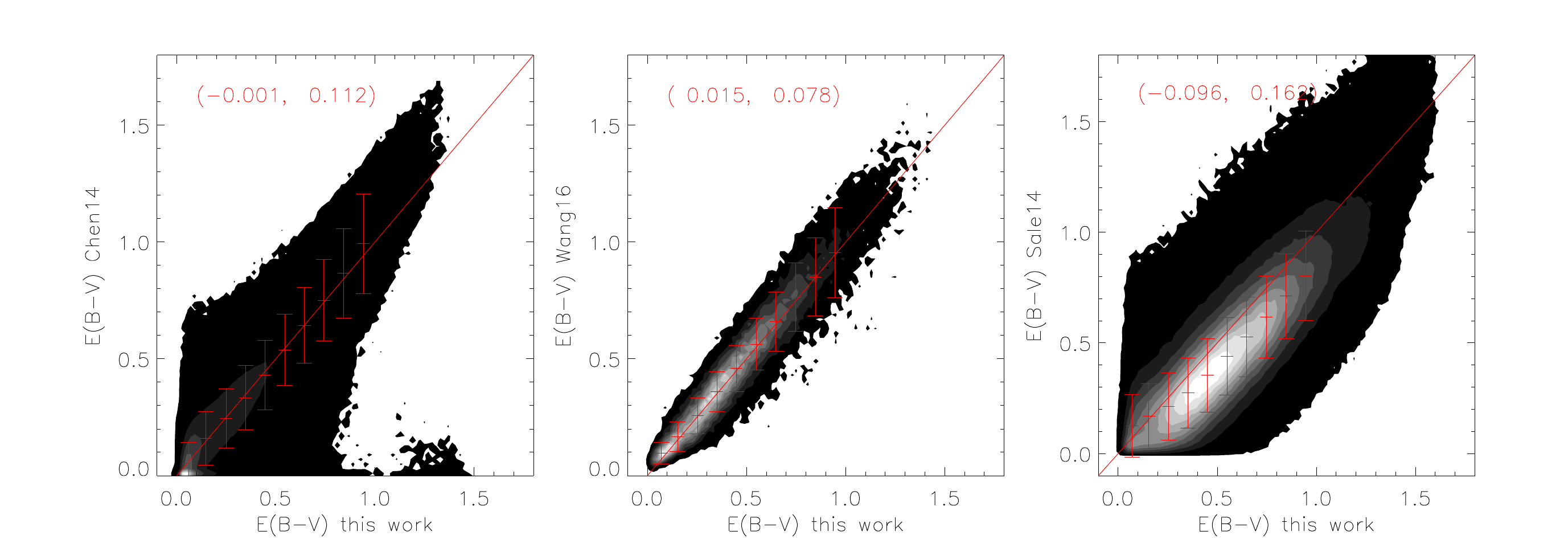}
  \caption{Comparisons of $E(B-V)$ values derived in the current work and those
deduced by \citet{Chen2014} (left), by \citet{Wang2016} (middle) and by
\citet{Sale2014} (right). Red pluses and the error bars represent
medians and standard deviations in the individual bins. Red straight lines denoting complete
equality are also overplotted to guide the eyes. Means and standard deviations of the 
differences (ours minus those from the literature), 
are marked in the individual panels.}
  \label{extc}
\end{figure*}

 \begin{figure*}
    \centering
  \includegraphics[width=0.98\textwidth]{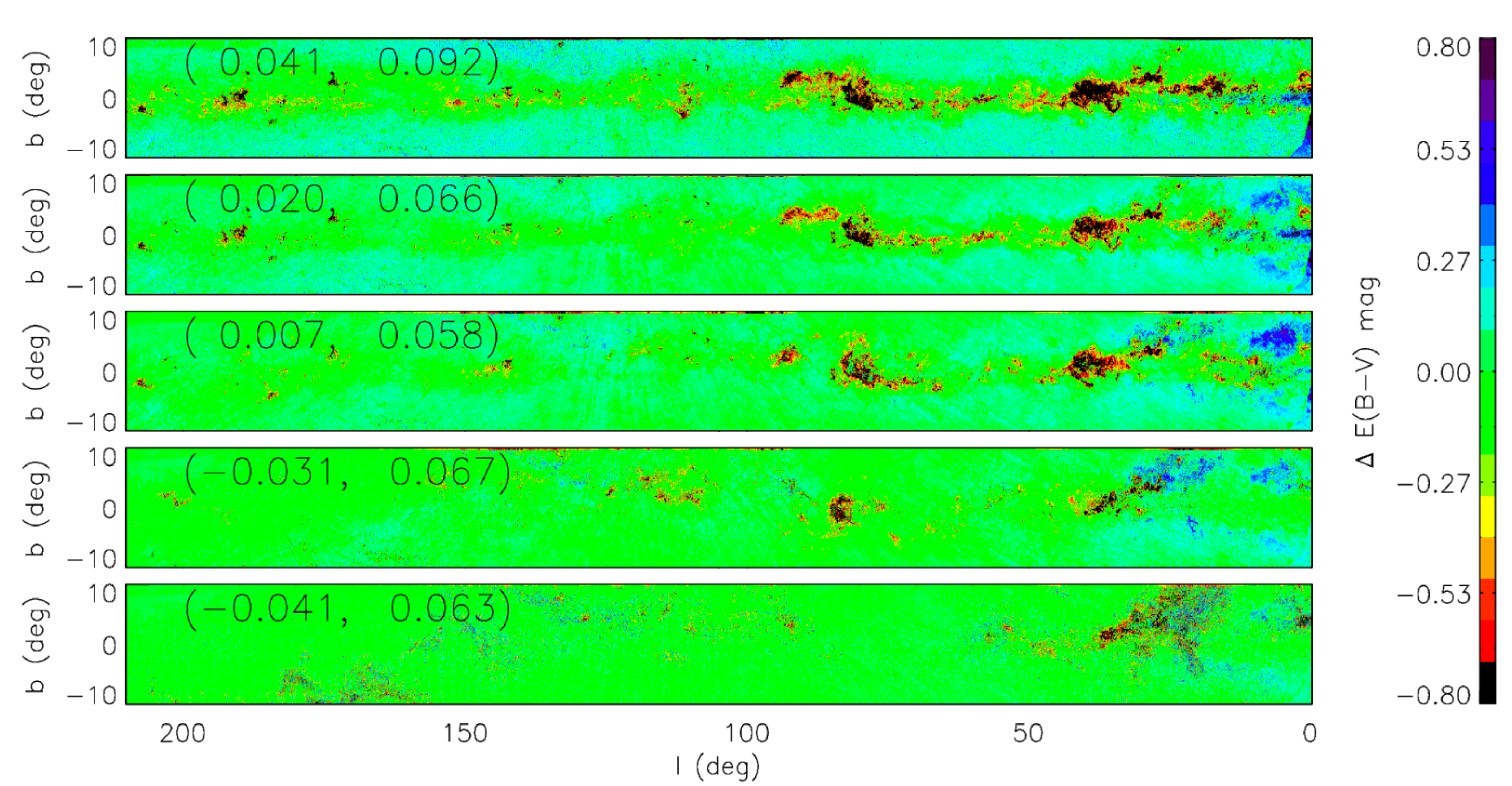}
  \caption{Comparison of our current 3D map with that of \citet{Green2018}. 
The panels show the differences of $E(B-V)$ (ours minus those of Green et al.), integrated to, from bottom to top, 
400, 800, 1600, 2800 and 5000 pc.
Means and standard deviations of the 
differences are marked in the individual panels.}
  \label{extcgreen}
\end{figure*}

We provide 3D maps for \egk, \ebr\ and \ehk\ separately. In order to derive line-of-sight extinction or colour excess in  
other passbands, one must assume certain relation between the extinction or colour excess in different bands, 
namely the extinction law. For example, the line-of-sight extinction in Gaia $G$-band, $A_G$, can be calculated from Eq.~(3)
 based on the extinction law of \citet{Yuan2013}.

Using our spectroscopic training data set, we are able to measure the reddening coefficients
for the Gaia and 2MASS photometric bands using a method similar to \citet{Yuan2013}. We select an ultra-high quality
spectroscopic sample from our final spectroscopic training sample with the criteria: 
spectral S/N  $>$ 50  and 
photometric uncertainties err($G$,  \gb,  \gr, $J, ~H,$ \ks, $W1$) $<$ 0.05\,mag. The requirements lead to a sample
consisting 1,447,606 stars. We then calculate values of $E(G_{\rm RP}-J)$ for the individual stars in this sample. 
Fig.~\ref{extl} plots $E(G-K_{\rm S})$, $E(G_{\rm RP}-J)$ and $E(H-K_{\rm S})$ versus \ebr\ for
all stars in this ultra-high quality spectroscopic sample.
We obtain colour excess ratios by linear fits of the median values and the origin.
The results are listed in Table~2.  Assuming that $R_J$ = 0.82, $R_H$ = 0.52 and $R_{K_{\rm S}}$ = 0.35 
as given by \citet{Yuan2013}, we obtain the extinction coefficients for the Gaia passbands. 
The results are also listed in Table~2. 
From the extinction coefficients, the line-of-sight extinction in 
\gb\ and \gr\ bands is then respectively given by,
\begin{equation}
A_{G_{\rm BP}} = 2.43 E(G_{\rm BP} - G_{\rm RP})
\end{equation}
and
\begin{equation}
A_{G_{\rm RP}} = 1.43 E(G_{\rm BP} - G_{\rm RP}).
\end{equation}

\subsection{Comparison with previous work}

To examine the accuracy of colour excesses derived in the current work with a machine learning algorithm, we compare 
our results with measurements from a number of other studies, including,

\begin{enumerate}
\item Values of $r$-band extinction of over 13 million stars in the Galactic anti-centre from \citet{Chen2014}, 
obtained by SED fitting to photometric measurements from the optical to the near-IR
($g,~r$,  and $i$ from the XSTPS-GAC, $J,~H$, and \ks\ from the 2MASS and $W1$, and $W2$ from the WISE).
\item Values of $K_{\rm S}$-band extinction of over 0.1 million stars observed by the APOGEE survey from \citet{Wang2016}, 
derived with a Bayesian approach by taking into account spectroscopic constraints 
from the APOGEE stellar parameters and photometric constraints from the 2MASS, as well as prior knowledge of
the Milky Way.
\item Values of monochromatic extinction at 5495\AA, $A_0$, of over 38 million stars in the Northern 
Galactic plane observed by IPHAS from \citet{Sale2014}, derived with a method based on a
hierarchical Bayesian model using the IPHAS photometry.
\end{enumerate}

Fig.~\ref{extc} compares our results with those from previous studies mentioned above. 
We cross-match our Gaia/2MASS/WISE sample with those of 
\citet{Chen2014}, \citet{Wang2016} and \citet{Sale2014} with a search radius of 1\,arcsec. 
For consistency, all the colour excess or extinction values
have been re-scaled to $E(B-V)$ using appropriate extinction
laws. We convert our current estimate of colour excess \ebr\ to $E(B-V)$ using the extinction coefficients 
presented in Sect.~5.1, and this yields $E(B-V)$ = 0.75 \ebr. The values of $A_r$ in \citet{Chen2014} are 
converted to those of $E(B-V)$ using the extinction law of \citet{Yuan2013}, which gives $E(B-V)$ = 0.43 $A_r$.
For $A_{K_{\rm S}}$ of \citet{Wang2016}, we use the \citet{Cardelli1989} extinction
law and have $E(B-V)$ = 2.77 $A_{K_{\rm S}}$. For $A_0$ from \citet{Sale2014}, we use the 
relation, $A_0$ = 1.003 $A_V$, from \citet{Sale2014} and assume $R_V$ = 3.1. This gives $E(B-V)$ = 0.32 $A_0$.

Fig.~\ref{extc} shows good agreement for all comparisons. Our
measurements, compared to those of \citet{Chen2014}, have an average
difference of only $-$0.001\,mag, along with an rms scatter of 0.11\,mag. 
The typical $r$-band extinction uncertainties for stars of photometric errors $\sim$ 0.05 mag 
in \citet{Chen2014} are about 0.16\,mag, corresponding to $\sim$ 0.07\,mag in $E(B-V)$. We can conclude
that our current measurements and those of Chen et al. contribute equally to the aforementioned dispersion, suggesting that
the current machine-learning method achieves an accuracy similar to the traditional SED fitting method.
There are a few stars have large $E(B-V)$ values from the current work ($\sim$ 1 -- 1.5\,mag), but small ones in
\citet{Chen2014}. Those stars are found to be very cool stars (spectral type M5-M6) that have intrinsic colour  
$(g-i)_0$ $\sim$ 3\,mag as found by \citet{Chen2014}.  
Due to the lack of very cool stars in our current spectroscopic training sample, we are not able to 
recover the colour excesses of those stars correctly. However, those stars contribute only 0.1\,per\,cent of the entire sample.
They do not have a significant effect on the 3D colour excess maps presented here.

Our results are well correlated with those of \citet{Wang2016}. The mean difference is only 0.015\,mag and the 
dispersion, 0.078\,mag, is the smallest among the comparisons. This is 
probably because the extinction values of \citet{Wang2016} are 
derived from spectroscopic data that have the smallest uncertainties. The dispersion is mainly contributed by
the uncertainties of our work, which is $\sim$ 0.07\,mag in $E(B-V)$.

Compared to \citet{Sale2014}, our results are systematical larger, with a mean
difference of $-$0.096\,mag, and a relatively large dispersion of 0.16\,mag. 
The offset and dispersion may partly be caused by the different data sets and methods used, as well as by the 
variations of extinction coefficients in the Galactic plane. 

Finally we compare our 3D colour excess map to the most recent 3D map 
of \citet{Green2018}. The latter covers the whole Northern sky ($\delta$ $\ge$ $-$30\degr)
out to a distance of several kpcs. The two maps overlap in the Galactic longitude range 0\degr\ $<$ $l$ $<$ 210\degr.
The map of 
Green et al. uses HEALPix pixelization scheme \citep{Gorski2005}. Depending on the regions, the pixel scale varies from 
 3.4 to 55\,arcmin. For each distance bin, we convert their map to the angular resolution of our map
 (6 $\times$ 6\,arcmin) by 2D linear interpolations.  We convert the maps of Green et al. to $E(B-V)$ using
 their relation, $E(B-V)$ = 0.996 $\times$ (Bayestar17). 
Fig.~\ref{extcgreen} shows the differences of cumulative values of $E(B-V)$ between our map and that of Green et al., 
integrated to distances 400, 800, 1600, 2800 and 5000\,pc, respectively.   
Overall, the two results agree well, with small differences of averages smaller than $\sim$ 0.04\,mag, and 
dispersions between 0.06 and 0.09\,mag. At larger distances, there are some regions ($l$ $\sim$ 20\degr $-$
80\degr and $b$ $\sim$ 0) showing large discrepancies, where our estimated colour excesses are systematically
smaller than those from \citet{Green2018}. This is mainly due to the small distance depths of our map in those regions
(see Fig.~\ref{dmax}).

\section{Summary}

The Gaia DR2 has provided us a great opportunity to study the 3D 
distribution of dust grains in the Galactic disk.
By combing the high quality optical photometry from the Gaia DR2  and those
from the 2MASS and WISE in the IR, we have simultaneously 
derived values of colour excess \egk, \ebr\ and \ehk\ for over 56 million stars, 
using the Random Forest regression, a machine learning algorithm.
In doing so, we have built an empirical training sample of stars selected  
from several large-scale spectroscopic surveys, including the LAMOST, SEGUE and
APOGEE. We derive values of colour excess \egk, \ebr\ and \ehk\ for over 3 million stars
in the spectroscopic sample with the star-pair technique. 
A comparison with results in the literature shows good agreement and that our current results have an
accuracy comparable to those derived from the
SED fitting method or Bayesian approaches, with typical uncertainties of about 0.07\,mag in $E(B-V)$.
However, the machine learning technique adopted in the current work is much faster than those traditional methods.
Values of \egk, \ebr\ and \ehk\ for the 56 million stars are available upon request by email.

By combining our colour excess values and the distances of \citet{Bailer2018} estimated from the 
Gaia parallaxes, we have constructed high-quality 3D colour excess map
for the entire Galactic plane (0\degr\ $<$ $l$ $<$ 360\degr and $|b|$ $<$ 10\degr).
 The map covers over 7000\,deg$^2$
at an angular resolution of 6\,arcmin, out to a distance of about 4 $-$ 6\,kpc from the Sun. The newly built map is in good agreement with 
those in the literature. 
The map will be public available, and
should be quite useful for follow up studies of the Milky Way disk.

Finally, using the spectroscopic sample we have calculated colour excess ratios and 
the extinction coefficients for the Gaia DR2 photometric bands. Empirically, we
have  $R_G$ = 2.50 $\pm$ 0.03,  $R_{G_{\rm BP}}$ = 3.24 $\pm$ 0.02 and $R_{G_{\rm RP}}$ = 1.91 $\pm$ 0.02.
The extinction coefficients can be used to convert our colour excesses to 
line-of-sight extinction in the Gaia DR2 bands (Eqs.~3, 5 and 6).

\section*{Acknowledgements}

We want to thank the referee, Prof. Coryn Bailer-Jones, for his insightful comments.
This work is partially supported by National Key Basic Research Program of China
2014CB845700 and  National
Natural Science Foundation of China 11803029, U1531244 and 11833006. HBY is supported by NSFC grant  
No.~11603002 and Beijing Normal University grant No.~310232102.
This research made use of the cross-match service provided by CDS, Strasbourg.

This work has made use of data products from the Guoshoujing Telescope (the
Large Sky Area Multi-Object Fibre Spectroscopic Telescope, LAMOST). LAMOST
is a National Major Scientific Project built by the Chinese Academy of
Sciences. Funding for the project has been provided by the National
Development and Reform Commission. LAMOST is operated and managed by the
National Astronomical Observatories, Chinese Academy of Sciences.

Funding for the Sloan Digital Sky Survey (SDSS) has been provided by the Alfred P. Sloan Foundation, the Participating Institutions, the National Aeronautics and Space Administration, the National Science Foundation, the U.S. Department of Energy, the Japanese Monbukagakusho, and the Max Planck Society. The SDSS Web site is http://www.sdss.org/.

The SDSS is managed by the Astrophysical Research Consortium (ARC) for the Participating Institutions. The Participating Institutions are The University of Chicago, Fermilab, the Institute for Advanced Study, the Japan Participation Group, The Johns Hopkins University, Los Alamos National Laboratory, the Max-Planck-Institute for Astronomy (MPIA), the Max-Planck-Institute for Astrophysics (MPA), New Mexico State University, University of Pittsburgh, Princeton University, the United States Naval Observatory, and the University of Washington.

This work presents results from the European Space Agency (ESA) space mission Gaia. Gaia data are being processed by the Gaia Data Processing and Analysis Consortium (DPAC). Funding for the DPAC is provided by national institutions, in particular the institutions participating in the Gaia MultiLateral Agreement (MLA). The Gaia mission website is https://www.cosmos.esa.int/gaia. The Gaia archive website is https://archives.esac.esa.int/gaia.

This publication makes use of data products from the Two Micron All Sky Survey, which is a joint project of the University of Massachusetts and the Infrared Processing and Analysis Center/California Institute of Technology, funded by the National Aeronautics and Space Administration and the National Science Foundation.

This publication makes use of data products from the Wide-field Infrared Survey Explorer, which is a joint project of the University of California, Los Angeles, and the Jet Propulsion Laboratory/California Institute of Technology, funded by the National Aeronautics and Space Administration.

\bibliographystyle{mn2e}
\bibliography{diskext}

\begin{thebibliography}{48}
\expandafter\ifx\csname natexlab\endcsname\relax\def\natexlab#1{#1}\fi

\bibitem[{{Andrae} {et~al.}(2018){Andrae}, {Fouesneau}, {Creevey}, {Ordenovic},
  {Mary}, {Burlacu}, {Chaoul}, {Jean-Antoine-Piccolo}, {Kordopatis}, {Korn},
  {Lebreton}, {Panem}, {Pichon}, {Th{\'e}venin}, {Walmsley}, \&
  {Bailer-Jones}}]{Andrae2018}
{Andrae}, R., {et~al.} 2018, \aap, 616, A8

\bibitem[{{Bailer-Jones}(2011)}]{Bailer2011}
{Bailer-Jones}, C.~A.~L. 2011, \mnras, 411, 435

\bibitem[{{Bailer-Jones} {et~al.}(2018){Bailer-Jones}, {Rybizki}, {Fouesneau},
  {Mantelet}, \& {Andrae}}]{Bailer2018}
{Bailer-Jones}, C.~A.~L., {Rybizki}, J., {Fouesneau}, M., {Mantelet}, G., \&
  {Andrae}, R. 2018, \aj, 156, 58

\bibitem[{{Berry} {et~al.}(2012){Berry}, {Ivezi{\'c}}, {Sesar}, {Juri{\'c}},
  {Schlafly}, {Bellovary}, {Finkbeiner}, {Vrbanec}, {Beers}, {Brooks},
  {Schneider}, {Gibson}, {Kimball}, {Jones}, {Yoachim}, {Krughoff}, {Connolly},
  {Loebman}, {Bond}, {Schlegel}, {Dalcanton}, {Yanny}, {Majewski}, {Knapp},
  {Gunn}, {Allyn Smith}, {Fukugita}, {Kent}, {Barentine}, {Krzesinski}, \&
  {Long}}]{Berry2012}
{Berry}, M., {et~al.} 2012, \apj, 757, 166

\bibitem[{{Cardelli} {et~al.}(1989){Cardelli}, {Clayton}, \&
  {Mathis}}]{Cardelli1989}
{Cardelli}, J.~A., {Clayton}, G.~C., \& {Mathis}, J.~S. 1989, \apj, 345, 245

\bibitem[{{Chambers} {et~al.}(2016){Chambers}, {Magnier}, {Metcalfe},
  {Flewelling}, {Huber}, {Waters}, {Denneau}, {Draper}, {Farrow}, {Finkbeiner},
  {Holmberg}, {Koppenhoefer}, {Price}, {Saglia}, {Schlafly}, {Smartt},
  {Sweeney}, {Wainscoat}, {Burgett}, {Grav}, {Heasley}, {Hodapp}, {Jedicke},
  {Kaiser}, {Kudritzki}, {Luppino}, {Lupton}, {Monet}, {Morgan}, {Onaka},
  {Stubbs}, {Tonry}, {Banados}, {Bell}, {Bender}, {Bernard}, {Botticella},
  {Casertano}, {Chastel}, {Chen}, {Chen}, {Cole}, {Deacon}, {Frenk},
  {Fitzsimmons}, {Gezari}, {Goessl}, {Goggia}, {Goldman}, {Grebel}, {Hambly},
  {Hasinger}, {Heavens}, {Heckman}, {Henderson}, {Henning}, {Holman}, {Hopp},
  {Ip}, {Isani}, {Keyes}, {Koekemoer}, {Kotak}, {Long}, {Lucey}, {Liu},
  {Martin}, {McLean}, {Morganson}, {Murphy}, {Nieto-Santisteban}, {Norberg},
  {Peacock}, {Pier}, {Postman}, {Primak}, {Rae}, {Rest}, {Riess}, {Riffeser},
  {Rix}, {Roser}, {Schilbach}, {Schultz}, {Scolnic}, {Szalay}, {Seitz},
  {Shiao}, {Small}, {Smith}, {Soderblom}, {Taylor}, {Thakar}, {Thiel},
  {Thilker}, {Urata}, {Valenti}, {Walter}, {Watters}, {Werner}, {White},
  {Wood-Vasey}, \& {Wyse}}]{Chambers2016}
{Chambers}, K.~C., {et~al.} 2016, ArXiv e-prints: 1612.05560

\bibitem[{{Chen} {et~al.}(2017{\natexlab{a}}){Chen}, {Liu}, {Ren}, {Yuan},
  {Huang}, {Yu}, {Xiang}, {Wang}, {Tian}, \& {Zhang}}]{Chen2017b}
{Chen}, B.-Q., {et~al.} 2017{\natexlab{a}}, \mnras, 472, 3924

\bibitem[{{Chen} {et~al.}(2017{\natexlab{b}}){Chen}, {Liu}, {Yuan}, {Robin},
  {Huang}, {Xiang}, {Wang}, {Ren}, {Tian}, \& {Zhang}}]{Chen2017a}
{Chen}, B.-Q., {et~al.} 2017{\natexlab{b}}, \mnras, 464, 2545

\bibitem[{{Chen} {et~al.}(2018){Chen}, {Liu}, {Yuan}, {Xiang}, {Huang}, {Wang},
  {Zhang}, \& {Tian}}]{Chen2018}
{Chen}, B.-Q., {Liu}, X.-W., {Yuan}, H.-B., {Xiang}, M.-S., {Huang}, Y.,
  {Wang}, C., {Zhang}, H.-W., \& {Tian}, Z.-J. 2018, \mnras, 476, 3278

\bibitem[{{Chen} {et~al.}(2014){Chen}, {Liu}, {Yuan}, {Zhang}, {Schultheis},
  {Jiang}, {Huang}, {Xiang}, {Zhao}, {Yao}, \& {Lu}}]{Chen2014}
{Chen}, B.-Q., {et~al.} 2014, \mnras, 443, 1192

\bibitem[{{Chen} {et~al.}(2013){Chen}, {Schultheis}, {Jiang}, {Gonzalez},
  {Robin}, {Rejkuba}, \& {Minniti}}]{Chen2013}
{Chen}, B.~Q., {Schultheis}, M., {Jiang}, B.~W., {Gonzalez}, O.~A., {Robin},
  A.~C., {Rejkuba}, M., \& {Minniti}, D. 2013, \aap, 550, A42

\bibitem[{{Churchwell} {et~al.}(2009){Churchwell}, {Babler}, {Meade},
  {Whitney}, {Benjamin}, {Indebetouw}, {Cyganowski}, {Robitaille}, {Povich},
  {Watson}, \& {Bracker}}]{Churchwell2009}
{Churchwell}, E., {et~al.} 2009, \pasp, 121, 213

\bibitem[{{Drew} {et~al.}(2005){Drew}, {Greimel}, {Irwin}, {Aungwerojwit},
  {Barlow}, {Corradi}, {Drake}, {G{\"a}nsicke}, {Groot}, {Hales}, {Hopewell},
  {Irwin}, {Knigge}, {Leisy}, {Lennon}, {Mampaso}, {Masheder}, {Matsuura},
  {Morales-Rueda}, {Morris}, {Parker}, {Phillipps}, {Rodriguez-Gil}, {Roelofs},
  {Skillen}, {Sokoloski}, {Steeghs}, {Unruh}, {Viironen}, {Vink}, {Walton},
  {Witham}, {Wright}, {Zijlstra}, \& {Zurita}}]{Drew2005}
{Drew}, J.~E., {et~al.} 2005, \mnras, 362, 753

\bibitem[{{Evans} {et~al.}(2018){Evans}, {Riello}, {De Angeli}, {Carrasco},
  {Montegriffo}, {Fabricius}, {Jordi}, {Palaversa}, {Diener}, {Busso},
  {Cacciari}, {van Leeuwen}, {Burgess}, {Davidson}, {Harrison}, {Hodgkin},
  {Pancino}, {Richards}, {Altavilla}, {Balaguer-N{\'u}{\~n}ez}, {Barstow},
  {Bellazzini}, {Brown}, {Castellani}, {Cocozza}, {De Luise}, {Delgado},
  {Ducourant}, {Galleti}, {Gilmore}, {Giuffrida}, {Holl}, {Kewley}, {Koposov},
  {Marinoni}, {Marrese}, {Osborne}, {Piersimoni}, {Portell}, {Pulone},
  {Ragaini}, {Sanna}, {Terrett}, {Walton}, {Wevers}, \&
  {Wyrzykowski}}]{Evans2018}
{Evans}, D.~W., {et~al.} 2018, \aap, 616, A4

\bibitem[{{Gaia Collaboration} {et~al.}(2018{\natexlab{a}}){Gaia
  Collaboration}, {Babusiaux}, {van Leeuwen}, {Barstow}, {Jordi}, {Vallenari},
  {Bossini}, {Bressan}, {Cantat-Gaudin}, {van Leeuwen}, \& et~al.}]{Gaia2018b}
{Gaia Collaboration}, {et~al.} 2018{\natexlab{a}}, \aap, 616, A10

\bibitem[{{Gaia Collaboration} {et~al.}(2018{\natexlab{b}}){Gaia
  Collaboration}, {Brown}, {Vallenari}, {Prusti}, {de Bruijne}, {Babusiaux},
  {Bailer-Jones}, {Biermann}, {Evans}, {Eyer}, \& et~al.}]{Gaia2018}
{Gaia Collaboration}, {et~al.} 2018{\natexlab{b}}, \aap, 616, A1

\bibitem[{{Gaia Collaboration} {et~al.}(2016){Gaia Collaboration}, {Prusti},
  {de Bruijne}, {Brown}, {Vallenari}, {Babusiaux}, {Bailer-Jones}, {Bastian},
  {Biermann}, {Evans}, \& et~al.}]{Gaia2016}
{Gaia Collaboration}, {et~al.} 2016, \aap, 595, A1

\bibitem[{{Gontcharov}(2017)}]{Gontcharov2017}
{Gontcharov}, G.~A. 2017, Astronomy Letters, 43, 472

\bibitem[{{G{\'o}rski} {et~al.}(2005){G{\'o}rski}, {Hivon}, {Banday},
  {Wandelt}, {Hansen}, {Reinecke}, \& {Bartelmann}}]{Gorski2005}
{G{\'o}rski}, K.~M., {Hivon}, E., {Banday}, A.~J., {Wandelt}, B.~D., {Hansen},
  F.~K., {Reinecke}, M., \& {Bartelmann}, M. 2005, \apj, 622, 759

\bibitem[{{Green} {et~al.}(2018){Green}, {Schlafly}, {Finkbeiner}, {Rix},
  {Martin}, {Burgett}, {Draper}, {Flewelling}, {Hodapp}, {Kaiser}, {Kudritzki},
  {Magnier}, {Metcalfe}, {Tonry}, {Wainscoat}, \& {Waters}}]{Green2018}
{Green}, G.~M., {et~al.} 2018, \mnras, 478, 651

\bibitem[{{Green} {et~al.}(2014){Green}, {Schlafly}, {Finkbeiner}, {Juri{\'c}},
  {Rix}, {Burgett}, {Chambers}, {Draper}, {Flewelling}, {Kudritzki}, {Magnier},
  {Martin}, {Metcalfe}, {Tonry}, {Wainscoat}, \& {Waters}}]{Green2014}
{Green}, G.~M., {et~al.} 2014, \apj, 783, 114

\bibitem[{{Green} {et~al.}(2015){Green}, {Schlafly}, {Finkbeiner}, {Rix},
  {Martin}, {Burgett}, {Draper}, {Flewelling}, {Hodapp}, {Kaiser}, {Kudritzki},
  {Magnier}, {Metcalfe}, {Price}, {Tonry}, \& {Wainscoat}}]{Green2015}
{Green}, G.~M., {et~al.} 2015, \apj, 810, 25

\bibitem[{{Hanson} {et~al.}(2016){Hanson}, {Bailer-Jones}, {Burgett},
  {Chambers}, {Hodapp}, {Kaiser}, {Tonry}, {Wainscoat}, \&
  {Waters}}]{Hanson2016}
{Hanson}, R.~J., {et~al.} 2016, \mnras, 463, 3604

\bibitem[{{Kirkpatrick} {et~al.}(2014){Kirkpatrick}, {Schneider},
  {Fajardo-Acosta}, {Gelino}, {Mace}, {Wright}, {Logsdon}, {McLean}, {Cushing},
  {Skrutskie}, {Eisenhardt}, {Stern}, {Balokovi{\'c}}, {Burgasser}, {Faherty},
  {Lansbury}, {Rich}, {Skrzypek}, {Fowler}, {Cutri}, {Masci}, {Conrow},
  {Grillmair}, {McCallon}, {Beichman}, \& {Marsh}}]{Kirkpatrick2014}
{Kirkpatrick}, J.~D., {et~al.} 2014, \apj, 783, 122

\bibitem[{{Lallement} {et~al.}(2018){Lallement}, {Capitanio}, {Ruiz-Dern},
  {Danielski}, {Babusiaux}, {Vergely}, {Elyajouri}, {Arenou}, \&
  {Leclerc}}]{Lallement2018}
{Lallement}, R., {et~al.} 2018, \aap, 616, A132

\bibitem[{{Lallement} {et~al.}(2014){Lallement}, {Vergely}, {Valette},
  {Puspitarini}, {Eyer}, \& {Casagrande}}]{Lallement2014}
{Lallement}, R., {Vergely}, J.-L., {Valette}, B., {Puspitarini}, L., {Eyer},
  L., \& {Casagrande}, L. 2014, \aap, 561, A91

\bibitem[{{Lindegren} {et~al.}(2018){Lindegren}, {Hern{\'a}ndez}, {Bombrun},
  {Klioner}, {Bastian}, {Ramos-Lerate}, {de Torres}, {Steidelm{\"u}ller},
  {Stephenson}, {Hobbs}, {Lammers}, {Biermann}, {Geyer}, {Hilger}, {Michalik},
  {Stampa}, {McMillan}, {Casta{\~n}eda}, {Clotet}, {Comoretto}, {Davidson},
  {Fabricius}, {Gracia}, {Hambly}, {Hutton}, {Mora}, {Portell}, {van Leeuwen},
  {Abbas}, {Abreu}, {Altmann}, {Andrei}, {Anglada}, {Balaguer-N{\'u}{\~n}ez},
  {Barache}, {Becciani}, {Bertone}, {Bianchi}, {Bouquillon}, {Bourda},
  {Br{\"u}semeister}, {Bucciarelli}, {Busonero}, {Buzzi}, {Cancelliere},
  {Carlucci}, {Charlot}, {Cheek}, {Crosta}, {Crowley}, {de Bruijne}, {de
  Felice}, {Drimmel}, {Esquej}, {Fienga}, {Fraile}, {Gai}, {Garralda},
  {Gonz{\'a}lez-Vidal}, {Guerra}, {Hauser}, {Hofmann}, {Holl}, {Jordan},
  {Lattanzi}, {Lenhardt}, {Liao}, {Licata}, {Lister}, {L{\"o}ffler},
  {Marchant}, {Martin-Fleitas}, {Messineo}, {Mignard}, {Morbidelli}, {Poggio},
  {Riva}, {Rowell}, {Salguero}, {Sarasso}, {Sciacca}, {Siddiqui}, {Smart},
  {Spagna}, {Steele}, {Taris}, {Torra}, {van Elteren}, {van Reeven}, \&
  {Vecchiato}}]{Lindegren2018}
{Lindegren}, L., {et~al.} 2018, \aap, 616, A2

\bibitem[{{Majewski} {et~al.}(2010){Majewski}, {Wilson}, {Hearty}, {Schiavon},
  \& {Skrutskie}}]{Majewski2010}
{Majewski}, S.~R., {Wilson}, J.~C., {Hearty}, F., {Schiavon}, R.~R., \&
  {Skrutskie}, M.~F. 2010, in IAU Symposium, Vol. 265, Chemical Abundances in
  the Universe: Connecting First Stars to Planets, ed. K.~{Cunha}, M.~{Spite},
  \& B.~{Barbuy}, 480--481

\bibitem[{{Marshall} {et~al.}(2006){Marshall}, {Robin}, {Reyl{\'e}},
  {Schultheis}, \& {Picaud}}]{Marshall2006}
{Marshall}, D.~J., {Robin}, A.~C., {Reyl{\'e}}, C., {Schultheis}, M., \&
  {Picaud}, S. 2006, \aap, 453, 635

\bibitem[{{Minniti} {et~al.}(2010){Minniti}, {Lucas}, {Emerson}, {Saito},
  {Hempel}, {Pietrukowicz}, {Ahumada}, {Alonso}, {Alonso-Garcia}, {Arias},
  {Bandyopadhyay}, {Barb{\'a}}, {Barbuy}, {Bedin}, {Bica}, {Borissova},
  {Bronfman}, {Carraro}, {Catelan}, {Clari{\'a}}, {Cross}, {de Grijs},
  {D{\'e}k{\'a}ny}, {Drew}, {Fari{\~n}a}, {Feinstein}, {Fern{\'a}ndez
  Laj{\'u}s}, {Gamen}, {Geisler}, {Gieren}, {Goldman}, {Gonzalez}, {Gunthardt},
  {Gurovich}, {Hambly}, {Irwin}, {Ivanov}, {Jord{\'a}n}, {Kerins}, {Kinemuchi},
  {Kurtev}, {L{\'o}pez-Corredoira}, {Maccarone}, {Masetti}, {Merlo},
  {Messineo}, {Mirabel}, {Monaco}, {Morelli}, {Padilla}, {Palma}, {Parisi},
  {Pignata}, {Rejkuba}, {Roman-Lopes}, {Sale}, {Schreiber}, {Schr{\"o}der},
  {Smith}, {}, {Soto}, {Tamura}, {Tappert}, {Thompson}, {Toledo}, {Zoccali}, \&
  {Pietrzynski}}]{Minniti2010}
{Minniti}, D., {et~al.} 2010, \na, 15, 433

\bibitem[{Pedregosa {et~al.}(2011)Pedregosa, Varoquaux, Gramfort, Michel,
  Thirion, Grisel, Blondel, Prettenhofer, Weiss, Dubourg, Vanderplas, Passos,
  Cournapeau, Brucher, Perrot, \& Duchesnay}]{scikit-learn}
Pedregosa, F., {et~al.} 2011, Journal of Machine Learning Research, 12, 2825

\bibitem[{{Planck Collaboration} {et~al.}(2014){Planck Collaboration},
  {Abergel}, {Ade}, {Aghanim}, {Alves}, {Aniano}, {Armitage-Caplan}, {Arnaud},
  {Ashdown}, {Atrio-Barandela}, \& et~al.}]{Planck2014}
{Planck Collaboration}, {et~al.} 2014, \aap, 571, A11

\bibitem[{{Reid} {et~al.}(2014){Reid}, {Menten}, {Brunthaler}, {Zheng}, {Dame},
  {Xu}, {Wu}, {Zhang}, {Sanna}, {Sato}, {Hachisuka}, {Choi}, {Immer},
  {Moscadelli}, {Rygl}, \& {Bartkiewicz}}]{Reid2014}
{Reid}, M.~J., {et~al.} 2014, \apj, 783, 130

\bibitem[{{Rezaei Kh.} {et~al.}(2017){Rezaei Kh.}, {Bailer-Jones}, {Hanson}, \&
  {Fouesneau}}]{Kh2017}
{Rezaei Kh.}, S., {Bailer-Jones}, C.~A.~L., {Hanson}, R.~J., \& {Fouesneau}, M.
  2017, \aap, 598, A125

\bibitem[{{Rezaei Kh.} {et~al.}(2018){Rezaei Kh.}, {Bailer-Jones}, {Schlafly},
  \& {Fouesneau}}]{Kh2018}
{Rezaei Kh.}, S., {Bailer-Jones}, C.~A.~L., {Schlafly}, E.~F., \& {Fouesneau},
  M. 2018, \aap, 616, A44

\bibitem[{{Robin} {et~al.}(2003){Robin}, {Reyl{\'e}}, {Derri{\`e}re}, \&
  {Picaud}}]{Robin2003}
{Robin}, A.~C., {Reyl{\'e}}, C., {Derri{\`e}re}, S., \& {Picaud}, S. 2003,
  \aap, 409, 523

\bibitem[{{Sale} {et~al.}(2014){Sale}, {Drew}, {Barentsen}, {Farnhill},
  {Raddi}, {Barlow}, {Eisl{\"o}ffel}, {Vink}, {Rodr{\'{\i}}guez-Gil}, \&
  {Wright}}]{Sale2014}
{Sale}, S.~E., {et~al.} 2014, \mnras, 443, 2907

\bibitem[{{Schlegel} {et~al.}(1998){Schlegel}, {Finkbeiner}, \&
  {Davis}}]{Schlegel1998}
{Schlegel}, D.~J., {Finkbeiner}, D.~P., \& {Davis}, M. 1998, \apj, 500, 525

\bibitem[{{Schultheis} {et~al.}(2014){Schultheis}, {Chen}, {Jiang}, {Gonzalez},
  {Enokiya}, {Fukui}, {Torii}, {Rejkuba}, \& {Minniti}}]{Schultheis2014}
{Schultheis}, M., {et~al.} 2014, \aap, 566, A120

\bibitem[{{Skrutskie} {et~al.}(2006){Skrutskie}, {Cutri}, {Stiening},
  {Weinberg}, {Schneider}, {Carpenter}, {Beichman}, {Capps}, {Chester},
  {Elias}, {Huchra}, {Liebert}, {Lonsdale}, {Monet}, {Price}, {Seitzer},
  {Jarrett}, {Kirkpatrick}, {Gizis}, {Howard}, {Evans}, {Fowler}, {Fullmer},
  {Hurt}, {Light}, {Kopan}, {Marsh}, {McCallon}, {Tam}, {Van Dyk}, \&
  {Wheelock}}]{Skrutskie2006}
{Skrutskie}, M.~F., {et~al.} 2006, \aj, 131, 1163

\bibitem[{{Wang} {et~al.}(2016){Wang}, {Shi}, {Pan}, {Chen}, {Zhao}, \&
  {Wicker}}]{Wang2016}
{Wang}, J., {Shi}, J., {Pan}, K., {Chen}, B., {Zhao}, Y., \& {Wicker}, J. 2016,
  \mnras, 460, 3179

\bibitem[{{Wright} {et~al.}(2010){Wright}, {Eisenhardt}, {Mainzer}, {Ressler},
  {Cutri}, {Jarrett}, {Kirkpatrick}, {Padgett}, {McMillan}, {Skrutskie},
  {Stanford}, {Cohen}, {Walker}, {Mather}, {Leisawitz}, {Gautier}, {McLean},
  {Benford}, {Lonsdale}, {Blain}, {Mendez}, {Irace}, {Duval}, {Liu}, {Royer},
  {Heinrichsen}, {Howard}, {Shannon}, {Kendall}, {Walsh}, {Larsen}, {Cardon},
  {Schick}, {Schwalm}, {Abid}, {Fabinsky}, {Naes}, \& {Tsai}}]{Wright2010}
{Wright}, E.~L., {et~al.} 2010, \aj, 140, 1868

\bibitem[{{Xiang} {et~al.}(2017){Xiang}, {Liu}, {Yuan}, {Huo}, {Huang}, {Wang},
  {Chen}, {Ren}, {Zhang}, {Tian}, {Yang}, {Shi}, {Zhao}, {Li}, {Zhao}, {Cui},
  {Li}, {Hou}, {Zhang}, {Zhang}, {Wang}, {Wu}, {Cao}, {Yan}, {Yan}, {Luo},
  {Zhang}, {Bai}, {Yuan}, {Dong}, {Lei}, \& {Li}}]{Xiang2017}
{Xiang}, M.-S., {et~al.} 2017, \mnras

\bibitem[{{Xu} {et~al.}(2018){Xu}, {Bian}, {Reid}, {Li}, {Zhang}, {Yan},
  {Dame}, {Menten}, {He}, {Liao}, \& {Tang}}]{Xu2018}
{Xu}, Y., {et~al.} 2018, \aap, 616, L15

\bibitem[{{Yanny} {et~al.}(2009){Yanny}, {Rockosi}, {Newberg}, {Knapp},
  {Adelman-McCarthy}, {Alcorn}, {Allam}, {Allende Prieto}, {An}, {Anderson},
  {Anderson}, {Bailer-Jones}, {Bastian}, {Beers}, {Bell}, {Belokurov},
  {Bizyaev}, {Blythe}, {Bochanski}, {Boroski}, {Brinchmann}, {Brinkmann},
  {Brewington}, {Carey}, {Cudworth}, {Evans}, {Evans}, {Gates}, {G{\"a}nsicke},
  {Gillespie}, {Gilmore}, {Nebot Gomez-Moran}, {Grebel}, {Greenwell}, {Gunn},
  {Jordan}, {Jordan}, {Harding}, {Harris}, {Hendry}, {Holder}, {Ivans},
  {Ivezi{\v c}}, {Jester}, {Johnson}, {Kent}, {Kleinman}, {Kniazev},
  {Krzesinski}, {Kron}, {Kuropatkin}, {Lebedeva}, {Lee}, {French Leger},
  {L{\'e}pine}, {Levine}, {Lin}, {Long}, {Loomis}, {Lupton}, {Malanushenko},
  {Malanushenko}, {Margon}, {Martinez-Delgado}, {McGehee}, {Monet}, {Morrison},
  {Munn}, {Neilsen}, {Nitta}, {Norris}, {Oravetz}, {Owen}, {Padmanabhan},
  {Pan}, {Peterson}, {Pier}, {Platson}, {Re Fiorentin}, {Richards}, {Rix},
  {Schlegel}, {Schneider}, {Schreiber}, {Schwope}, {Sibley}, {Simmons},
  {Snedden}, {Allyn Smith}, {Stark}, {Stauffer}, {Steinmetz}, {Stoughton},
  {SubbaRao}, {Szalay}, {Szkody}, {Thakar}, {Sivarani}, {Tucker}, {Uomoto},
  {Vanden Berk}, {Vidrih}, {Wadadekar}, {Watters}, {Wilhelm}, {Wyse}, {Yarger},
  \& {Zucker}}]{Yanny2009}
{Yanny}, B., {et~al.} 2009, \aj, 137, 4377

\bibitem[{{York} {et~al.}(2000){York}, {Adelman}, {Anderson}, {Anderson},
  {Annis}, {Bahcall}, {Bakken}, {Barkhouser}, {Bastian}, {Berman}, {Boroski},
  {Bracker}, {Briegel}, {Briggs}, {Brinkmann}, {Brunner}, {Burles}, {Carey},
  {Carr}, {Castander}, {Chen}, {Colestock}, {Connolly}, {Crocker}, {Csabai},
  {Czarapata}, {Davis}, {Doi}, {Dombeck}, {Eisenstein}, {Ellman}, {Elms},
  {Evans}, {Fan}, {Federwitz}, {Fiscelli}, {Friedman}, {Frieman}, {Fukugita},
  {Gillespie}, {Gunn}, {Gurbani}, {de Haas}, {Haldeman}, {Harris}, {Hayes},
  {Heckman}, {Hennessy}, {Hindsley}, {Holm}, {Holmgren}, {Huang}, {Hull},
  {Husby}, {Ichikawa}, {Ichikawa}, {Ivezi{\'c}}, {Kent}, {Kim}, {Kinney},
  {Klaene}, {Kleinman}, {Kleinman}, {Knapp}, {Korienek}, {Kron}, {Kunszt},
  {Lamb}, {Lee}, {Leger}, {Limmongkol}, {Lindenmeyer}, {Long}, {Loomis},
  {Loveday}, {Lucinio}, {Lupton}, {MacKinnon}, {Mannery}, {Mantsch}, {Margon},
  {McGehee}, {McKay}, {Meiksin}, {Merelli}, {Monet}, {Munn}, {Narayanan},
  {Nash}, {Neilsen}, {Neswold}, {Newberg}, {Nichol}, {Nicinski}, {Nonino},
  {Okada}, {Okamura}, {Ostriker}, {Owen}, {Pauls}, {Peoples}, {Peterson},
  {Petravick}, {Pier}, {Pope}, {Pordes}, {Prosapio}, {Rechenmacher}, {Quinn},
  {Richards}, {Richmond}, {Rivetta}, {Rockosi}, {Ruthmansdorfer}, {Sandford},
  {Schlegel}, {Schneider}, {Sekiguchi}, {Sergey}, {Shimasaku}, {Siegmund},
  {Smee}, {Smith}, {Snedden}, {Stone}, {Stoughton}, {Strauss}, {Stubbs},
  {SubbaRao}, {Szalay}, {Szapudi}, {Szokoly}, {Thakar}, {Tremonti}, {Tucker},
  {Uomoto}, {Vanden Berk}, {Vogeley}, {Waddell}, {Wang}, {Watanabe},
  {Weinberg}, {Yanny}, {Yasuda}, \& {SDSS Collaboration}}]{York2000}
{York}, D.~G., {et~al.} 2000, \aj, 120, 1579

\bibitem[{{Yuan} {et~al.}(2015){Yuan}, {Liu}, {Huo}, {Xiang}, {Huang}, {Chen},
  {Zhang}, {Sun}, {Wang}, {Zhang}, {Zhao}, {Luo}, {Shi}, {Li}, {Yuan}, {Dong},
  {Li}, {Hou}, \& {Zhang}}]{Yuan2015}
{Yuan}, H.-B., {et~al.} 2015, \mnras, 448, 855

\bibitem[{{Yuan} {et~al.}(2013){Yuan}, {Liu}, \& {Xiang}}]{Yuan2013}
{Yuan}, H.~B., {Liu}, X.~W., \& {Xiang}, M.~S. 2013, \mnras, 430, 2188

\end{thebibliography}






\label{lastpage}
\end{document}